# The First IRAM/PdBI Polarimetric Millimeter Survey of Active Galactic Nuclei⋆

## I. Global Properties of the Sample

S. Trippe, R. Neri, M. Krips, A. Castro-Carrizo, M. Bremer, V. Piétu, and A.L. Fontana

Institut de Radioastronomie Millimétrique (IRAM), 300 rue de la Piscine, F-38406 Saint Martin d'Hères, France

**Abstract.** We have studied the linear polarization of 86 active galactic nuclei (AGN) in the observed frequency range 80–267 GHz (3.7–1.1mm in wavelength), corresponding to *rest-frame* frequencies 82–738 GHz, with the IRAM Plateau de Bure Interferometer (PdBI). The large number of measurements, 441, makes our analysis the largest polarimetric AGN survey in this frequency range to date. We extracted polarization parameters via earth rotation polarimetry with unprecedented median precisions of ∼0.1% in polarization fractions and ∼1.2° in polarization angles. For 73 of 86 sources we detect polarization at least once. The degrees of polarization are as high as ∼19%, with the median over all sources being ∼4%. Source fluxes and polarizations are typically highly variable, with fractional variabilities up to ∼60%. We find that BLLac sources have on average the highest level of polarization. There appears to be no correlation between degree of polarization and redshift, indicating that there has been no substantial change of polarization properties since $z \simeq 2.4$. Our polarization and spectral index distributions are in good agreement with results found from various samples observed at cm/radio wavelengths; thus our frequency range is likely tracing the signature of synchrotron radiation without noticeable contributions from other emission mechanisms. The "millimeter-break" located at frequencies ≳1 THz appears to be not detectable in the frequency range covered by our survey.

## 1. Introduction

Active galactic nuclei (AGN) have been studied extensively in the wavelength range from cm-radio to $\gamma$ radiation in the last decades (see, e.g., Kembhavi & Narlikar 1999, or Krolik 1999, and references therein for a review). There is overwhelming observational evidence that the main source of their emission is accretion onto supermassive black holes (SMBH) with masses $M_\bullet \simeq 10^{6...9} M_\odot$ (e.g., Ferrarese & Ford 2005, and references therein). However, the properties of AGN at millimeter wavelengths are comparatively poorly known. Studies covering substantial (more than about ten) numbers of targets in this wavelength range mainly aim at fluxes or morphologies (e.g. Steppe et al. 1988, 1992, 1993; Teräsranta et al. 1998; Sadler et al. 2008; Hardcastle & Looney 2008).

The linear polarization of AGN contains important information on the physics of galactic nuclei. Polarization fractions and angles provide details on synchrotron emission, the geometry of emission regions, strength and orientation of magnetic fields, and (via Faraday rotation and/or depolarization) on particle densities and matter distributions of the surrounding or outflowing matter (see, e.g., Saikia & Salter 1988, and references therein). This makes a detailed overview on the polarization properties of AGN in the millimeter wavelength range highly valuable. However, existing studies are usually limited to few selected sources (e.g. Hobbs et al. 1978; Rudnick et al. 1978; Stevens et al. 1996, 1998; Attridge 2001), the work of Nartallo et al. (1998), which covers 26 sources, being a notable exception. One might also count similar studies of the black hole in the center of the Milky Way, Sgr A* (e.g. Marrone et al. 2006), and other local low-luminosity AGN (LLAGN; e.g. Bower, Falcke & Mellon 2002) in this context.

Even if sources are not resolved spatially, polarimetric analyses provide decisive information in combination with high-resolution maps (e.g. Very Long Baseline Interferometry). One example is the relation between position angles of AGN jets and polarization angles (e.g. Stevens et al. 1996, 1998; Nartallo et al. 1998). Those studies are able to constrain the magnetic field configurations in the jets and from this the types and geometries of shocks propagating through the jets.

Knowledge on the millimeter polarization of AGN is also important for technical reasons. All coherent receiver systems are sensitive to polarization by construction (see, e.g., Rohlfs & Wilson 2006 and references therein for an overview). As AGN are commonly used for flux or amplitude calibration of

*Send offprint requests to*: S. Trippe, e-mail: trippe@iram.fr

⋆ This study is based on observations carried out with the IRAM Plateau de Bure Interferometer. IRAM is supported by INSU/CNRS (France), MPG (Germany), and IGN (Spain).

radioastronomical data, a non-recognized polarization of their fluxes can introduce systematic calibration errors. Therefore the construction of a polarimetric AGN catalog is helpful for improving the quality of (millimeter) radioastronomical data. Additionally, such a catalog is needful for the calibration of polarimetric observations.

Like other radioastronomical observatories, the IRAM Plateau de Bure Interferometer (PdBI) uses AGN as phase and amplitude calibrators. We have used this fact to conduct a polarimetric survey of 86 sources observed from January 2007 to December 2009. The scope of this survey is to provide for the first time an overview on the polarimetric properties of AGN in the mm/radio domain based on a large number of targets. We intend to present our results in two papers: the present paper I provides a global overview on the survey, its methods and database, presents the results derived from analyses of the sample as whole, and puts the results into the context of AGN properties at cm/radio wavelengths. Paper II is going to provide a deep analysis of a few AGN from our sample for which multiple (more than ten), time- and frequency-resolved measurements are at hand – i.e. sources for which a statistical and time series analysis per object appears promising.

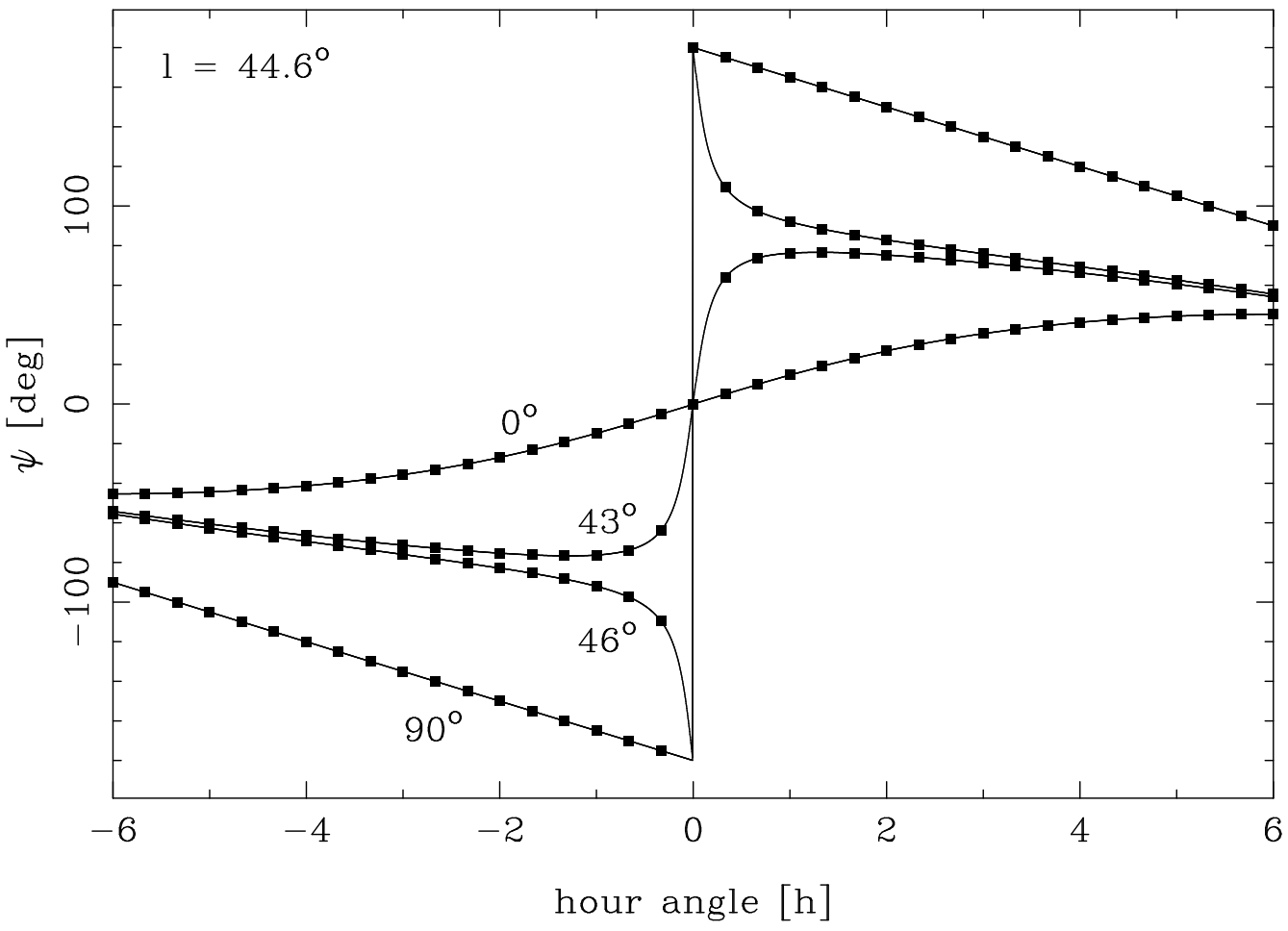

**Fig. 2.** Parallactic angle $\psi$ vs. hour angle as function of target declination $\delta$. The continuous lines mark the $\psi$ coverage of four targets located at declinations 0°, 43°, 46°, and 90°, respectively (see graph labels). The points mark the actual sampling typically obtained by PdBI observations of calibrators, i.e. one observation every ∼20 min. If $\delta > l = 44.6°$, the coverage becomes discontinuous at the transit.

## 2. Observations and Analysis

### 2.1. Earth Rotation Polarimetry

In January 2007 (January 2008 for the 2mm-band), the antennas of the IRAM Plateau de Bure Interferometer (PdBI) were equipped with dual linear polarization Cassegrain focus receivers. Since then, it is possible to observe both orthogonal polarizations – "horizontal" (H) and "vertical" (V) with respect to the antenna frame – simultaneously. Observations can be carried out non-simultaneous in three atmospheric windows located around wavelengths of 1.3mm, 2mm, and 3mm. Each of these bands covers a continuous range of frequencies; these ranges are 201–267 GHz for the 1.3mm band, 129–174 GHz for the 2mm band, and 80–116 GHz for the 3mm band. Within a given band, any frequency is available for observations (Winters & Neri 2008)[1].

The PdBI is not (yet) equipped for observations of all Stokes parameters. We collect linear polarization data on point sources via earth rotation polarimetry, i.e. we monitor the fluxes in the H and V channels as functions of parallactic angle $\psi$. Initially, all data ($H(\psi)$, $V(\psi)$) are recorded as function of hour angle $h$ (i.e. $H(h)$, $V(h)$). Before calculating polarization parameters, we therefore convert hour angles $h$ into parallactic angles $\psi$ via

$$\psi = \mathrm{atan2}[\sin(h), \tan(l)\cos(\delta) - \sin(\delta)\cos(h)] \quad (1)$$

with $l$ being the observatory latitude (for the PdBI, $l = 44.6°$), $\delta$ being the source declination, and atan2 being the quadrant-preserving arctangent.

For assessing the polarization of a source, we calculate the parameter

$$q(\psi) = \frac{V - H}{V + H}(\psi) \quad (2)$$

from the fluxes $H(\psi)$ and $V(\psi)$. As $H$ and $V$ correspond to the two parallel-hand products of the electric field vectors, the parameter $q(\psi)$ corresponds to a combination of the Stokes parameters $I$, $Q$, and $U$ as function of $\psi$:

$$q(\psi) \equiv \frac{Q}{I}\cos(2\psi) + \frac{U}{I}\sin(2\psi) \quad (3)$$

This means that $q(\psi)$ provides full information on linear polarization (see, e.g., Sault, Hamaker & Bregman 1996; Thompson, Moran & Swenson 2001).

For each PdBI project, we calculate $q(\psi)$ before performing a flux or amplitude calibration; this allows estimating the linear polarization of calibrator sources before using them for this purpose. As $q(\psi)$ is a relative parameter, the gain factors in numerator and denominator cancel out, meaning that a correct evaluation of $q(\psi)$ does not require a gain or flux calibration. This also means that a polarization analysis using $q(\psi)$ as parameter is not limited in accuracy by these calibration steps.

Due to the nature of polarized light and the fact that the PdBI antenna receivers are located in the Cassegrain foci, observing a polarized target results in $q(\psi)$ being a cosinusoidal signal with a period of 180°. The functional form of $q(\psi)$ is thus

$$q(\psi) \equiv m_L \cos[2(\psi - \chi)] + o \quad (4)$$

Here $m_L$ is the fraction of linear polarization (ranging from 0 to 1; in the following, we will express $m_L$ in units of %), $\chi$ is the polarization angle (ranging from 0° to 180° and counted from north to east), and $o$ is a global offset (we discuss the meaning of $o$ below). We note that Beltrán et al. (2004) used a very similar ansatz for deriving the polarization parameters of the BLLac object 1823+568 from PdBI calibration data.

For each dataset, we calculate $m_L$, $\chi$, and $o$ by means of a least-squares fit. This calculation is antenna-based, i.e. executed for each antenna separately. The final result is the average of the $N$ ($N \leq 6$) individual antenna results. Statistical errors are standard errors $\sigma_X/\sqrt{N}$ with $\sigma_X$ being the standard deviation of parameter $X$ over all antennas. We show an example for our analysis scheme in Fig. 1.

[1] http://www.iram.fr/IRAMFR/GILDAS/doc/pdf/pdbi-intro.pdf

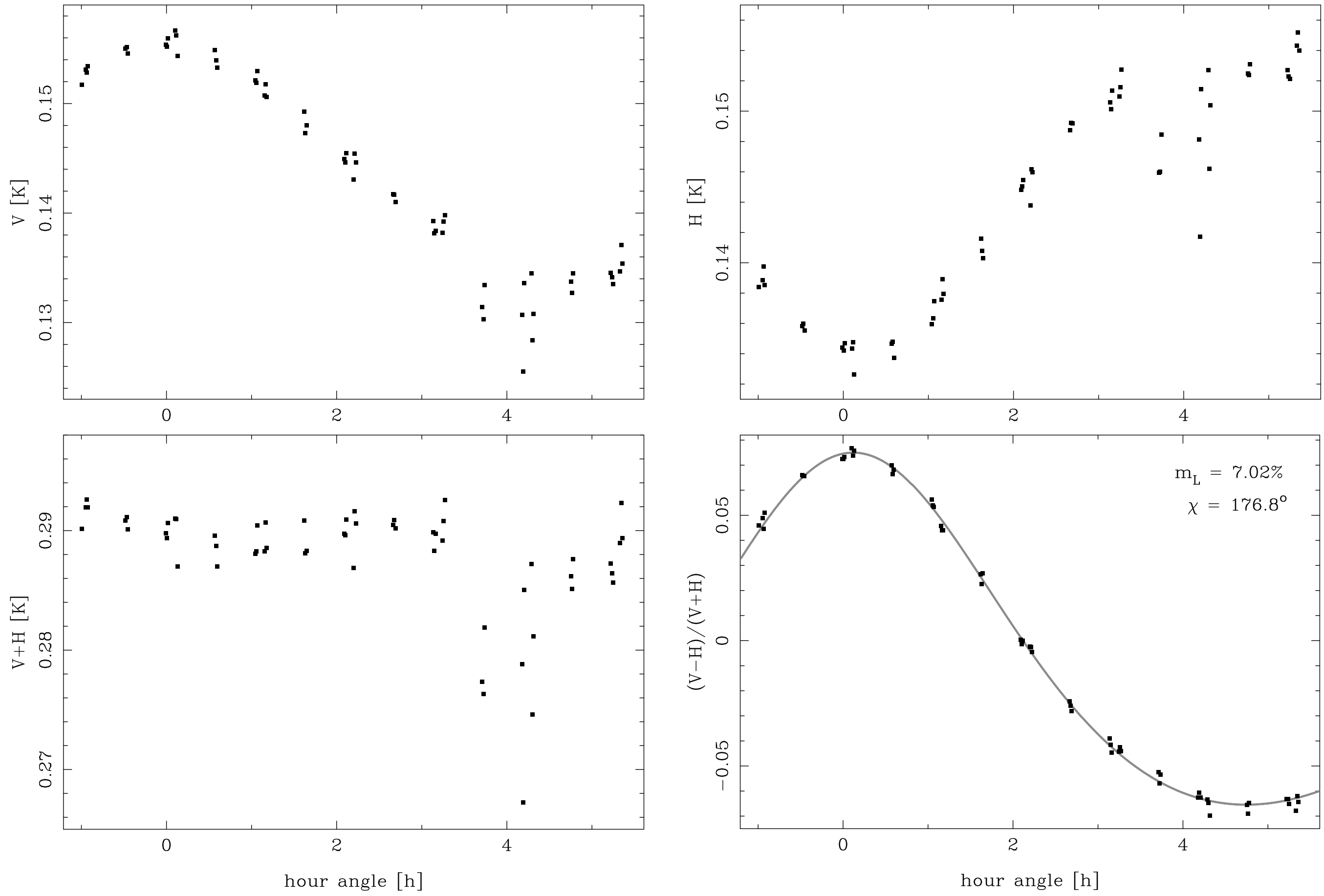

**Fig. 1.** A polarimetric analysis of the quasar 1642+690 as observed on May 3, 2009, at 103 GHz under good atmospheric conditions. This diagram illustrates our data and our analysis scheme using the results from one antenna as example. Analogous analyses are carried out for the other five antennas. We show here $V$ (top left), $H$ (top right), $(V+H)$ (bottom left), and $q \equiv (V-H)/(V+H)$ (bottom right) as functions of hour angle $h$. Please note the variations in the amplitude axes scales. Points are data, the continuous grey line in the bottom right panel corresponds to the best cosine fit to $q(\psi)$ with $\psi$ being the parallactic angle; please see Sect. 2.1 for a detailed explanation. As gain variations cancel out in $q(h)$, the $q(h)$ curve is smoother than the ones for $V(h)$, $H(h)$, and $(V+H)$; this is straightforward to see at $h \gtrsim 4$ when flux loss due to clouds occured. For the measurement shown here we find a degree of polarization $m_L = 7.02\%$ and a polarization angle $\chi = 176.8°$. From averaging the results from all six antennas, we find $m_L = 7.01 \pm 0.02\%$ and $\chi = 176.6 \pm 0.1°$ (statistical errors).

As one can see from Eq. 4, reliably extracting polarization parameters requires a proper sampling and coverage in $\psi$. We therefore accepted only measurements which (a) cover at least 4.5h in hour angle, (b) cover at least 67.5° (i.e. $3\pi/8$) in parallactic angle, and (c) are sampled with at least five observation per hour. A minimum angle coverage of 67.5° turned out to be a good choice as it covers a large enough part of the $q(\psi)$ curve to prevent the appearence of ambiguous solutions even in noisy data. Using the two redundant criteria (a) and (b) is necessary because of the behaviour of the parallactic angle for targets with declinations $\delta$ approximately equal or larger than the observatory latitude $l$; this we illustrate in Fig. 2. If $\delta > l$, the $\psi$ curve becomes discontinuous at transit. In those cases we use the coverage in $|\psi|$ (i.e. the absolute value of $\psi$) as selection criterion. If $\delta < l$ and $\delta \simeq l$, the $\psi$ coverage is continuous at transit but the angle swings quickly (within about 1h) from large negative to large positive values; relying on the $\psi$ coverage alone can thus lead to a poor sampling around the transit. Criterion (a) ensures that in any case a reasonable minimum number of data points (>10) is included into the analysis, thus granting a sufficient sampling.

In order to stabilize the model fitting, we iteratively rejected outlier data points with unusually strong (>33% w.r.t. the median) gain variations (caused, e.g., by unrecognized antenna shadowing) or strong (>0.1 w.r.t. the best fit) deviation (e.g. due to unrecognized technical problems) from the $q(\psi)$ curve.

## 2.2. Selection of Targets

Our targets are PdBI calibration sources. During each standard PdBI observation, one or two (spatially not resolved) AGN are monitored in parallel to the primary science target(s) for the purpose of phase and amplitude calibration. The AGN observations are thus a by-product of the standard interferometer operation; this makes it possible to conduct a large AGN survey without the need for dedicated observing time. Each calibra-tor is observed every ~20 min for ~2 min (usually 3 scans of 45 sec each) per observation. A typical PdBI observation lasts for several hours (~4–8h per science target), resulting in a corresponding number of calibrator measurements and coverage in $\psi$.

The selection of PdBI calibrators is based on the surveys by Steppe et al. (1988, 1992, 1993), Patnaik et al. (1992), Browne et al. (1998), and Wilkinson et al. (1998). The principal selection criterion is the flux density at 90 GHz which should be higher than 0.3 Jy. Due to source flux variability, this cutoff is actually a soft limit. Our list of potential targets thus includes all known mm/radio sources with minimum flux densities $\gtrsim$0.2 Jy at 90 GHz. Another constraint is the latidude of the PdBI ($l = 44.6°$) which excludes reasonable observations of targets with declinations $\delta \lesssim -20°$.

An AGN is actually observed if it can serve as a calibrator for a primary PdBI target. Therefore our sample is statistically incomplete; there is a bias towards bright sources located close to primary PdBI targets proposed by the astronomical community. This is important when comparing our results to other studies: On the one hand, our survey does not permit straightforward conclusions on physical parameters related to spatial distributions or luminosity functions of AGN samples. On the other hand, our method does not (at least not explicitly) impose a selection with respect to linear polarization, redshift, or spectral properties. Those parameters are therefore available for analysis.

### 2.3. Systematic Uncertainties

We discuss the *statistical* accuracies of our analysis results in Sect. 3.1. Identifying possible *systematic* errors is less straightforward. As given in Eq. 4, $q(\psi)$ is a relative parameter and therefore not affected by inaccuracies of flux or gain calibrations. However, this holds only for the case that the gains (efficiencies) of the channels H and V are identical. Systematic gain differences may occur for the following reasons:

- The two channels have intrinsically different sensitivities or efficiencies on timescales of a few hours or longer. Another possibility is a pointing offset on sky between the two channels. However, these properties were tested during the commissioning of the PdBI dual polarization receivers; their contribution was considered to be negligible (the IRAM commissioning team, *priv. comm.*).
- Instrumental polarization which is fixed with respect to the antenna frame. This effect we indeed may expect for the PdBI antennas as the receivers are located in the Cassegrain foci; this means that the entire optical system is aligned along the optical axis without rotating elements. Instrumental polarization would thus express itself as channel efficiencies that are different for the two polarizations, but do not vary with time or parallactic angle.

If this is the case, one can rewrite the H channel flux as $H(\psi) \rightarrow rH(\psi)$ with $r > 0$ being the ratio of the gains of channels H and V. Any $r \neq 1$ modifies $q(\psi)$ such that (1) the $q(\psi)$ curve experiences a small deformation which is symmetric, (2) $o \neq 0$ in Eq. 4, and (3) the polarization amplitude $m_L$ is slightly reduced. In order to take this effect into account, we introduced $o$ as parameter into our model (Eq. 4). From our data, we find $|o| = 0...0.07$, corresponding to $r = 0.85...1.2$. As one can easily evaluate numerically, such gain ratios can lead to a relative reduction of degrees of polarization $m_L$ by 0.5% at most. This means that for a source with a physical $m_L = 10\%$ we would actually measure $m'_L = 9.95\%$. Even when taking into account our good statistical accuracies (see Sect. 3.1), we can neglect this small effect.

Although the influence of instrumental polarization is most probably negligible – as long as it is not a function of $h$ or $\psi$ – we quantified the receiver polarization via a dedicated laboratory experiment. For this we made use of a spare receiver based at the IRAM headquarters in Grenoble (France); this receiver is identical to the ones mounted at the PdBI antennas. The receiver is exposed to an artificial radio signal which is perfectly linearly polarized by using a wire grid polariser. We compare recorded signal and emitted radiation in order to derive the instrumental polarization of the receiver system. With our laboratory setup we are able to test the frequency range 84–116 GHz. For this range we find receiver polarizations of ~0.3%. We thus can conclude that the influence of instrumental polarization (at least the component intrinsic to the receiver system) is indeed negligible for our study.

In order to assess the global systematic uncertainty of our analysis, we use the lowest reliable (i.e. resulting from measurements passing all automatic quality checks plus inspection by eye) $3\sigma$ upper limits $s_m$ on $m_L$ from our survey as references. The corresponding figures are

$s_m \lesssim 1.1\%$ for the 1.3mm band,
$s_m \lesssim 1.1\%$ for the 2mm band,
$s_m \lesssim 0.4\%$ for the 3mm band.

The true values are probably smaller for two reasons: (a) observations at higher frequencies suffer from lower signal-to noise ratios (S/N) in general, resulting in higher upper limits on $m_L$, and (b) low-number statistics. Both effects seem to affect the results we quote above: the numbers for the 1.3mm and 2mm bands are clearly larger than the one for the 3mm band. First, higher frequencies approximately correspond to higher upper limits as expected from effect (a). Second, we actually find for the 2mm band about the same value as for the 1.3mm band – as expected from effect (b) given that both samples are quite small (36 measurements each). We thus assume that the limits we quote here are actually conservative estimates – but that these are the numbers we have to use as long as additional information is not available.

## 3. Results

### 3.1. Data and Accuracies

We included all PdBI calibration source data obtained from January 2007 to December 2009 into our analysis. Using the procedure and selection criteria outlined in Section 2, we collected 441 polarization measurements for 86 AGN, resolved in time and frequency. Our analysis covers the *observed* frequency range 80–267 GHz (3.7–1.1mm in wavelength). For reasons of consistent bookkeeping, we use for all sources the epoch B1950 nomenclature throughout this article; in Table 2 we give commonly used alternative names for the most important targets. We present an overview over all observations in Table 2.

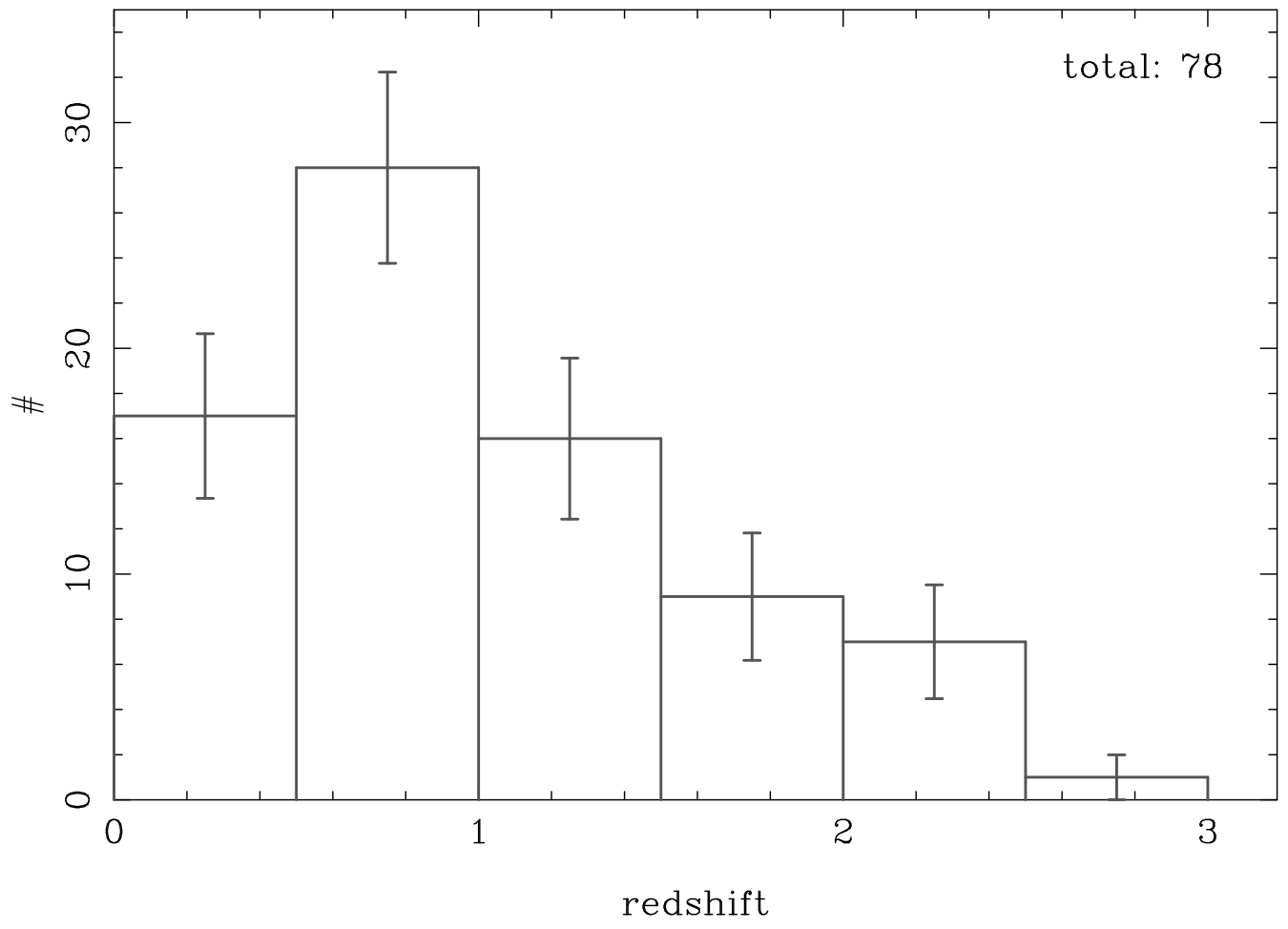

**Fig. 3.** The redshift distribution of our AGN sample. For 78 out of our 86 sources redshifts $z$ are available from the literature. Our sample covers redshifts from 0.02 to 2.69.

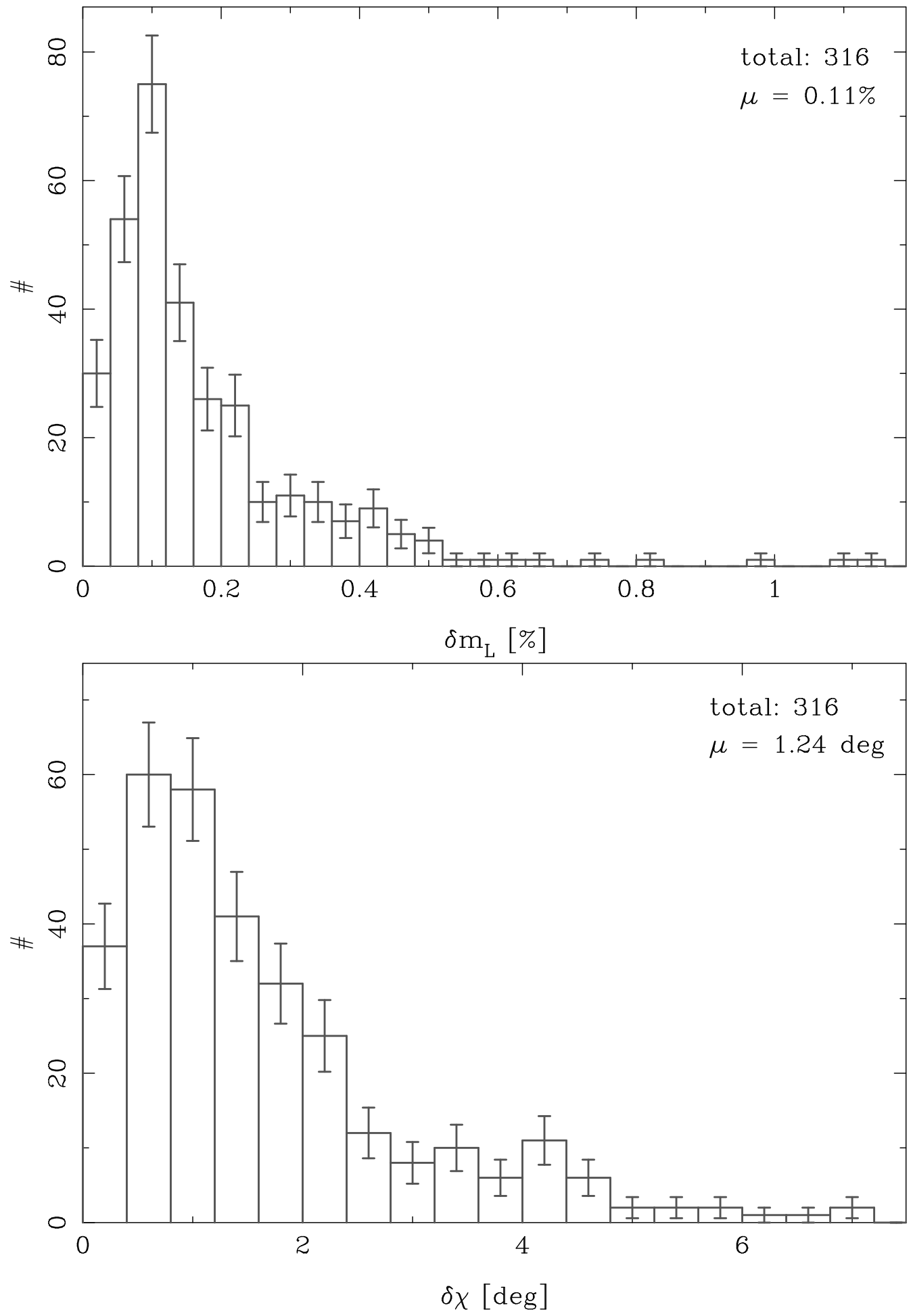

**Fig. 4.** Distributions of statistical errors in polarization fraction $\delta m_L$ (top panel) and polarization angles $\delta\chi$ (bottom panel) for all 273 measurements with detected linear polarization. Errorbars are binomial errors. The distribution of the $\delta p$ has a median $\mu \simeq$0.1% with the tail ranging out to ∼1%. The histogram of the $\delta\chi$ has $\mu \simeq 1.2°$ with the tail ranging out to $\simeq$ 7°. These values correspond to unprecedented precisions.

For 78 out of our 86 AGN, redshifts $z$ are available from the literature. We collected these figures from the NASA/IPAC Extragalactic Database (NED)[2] and the MOJAVE Survey Data Archive[3] (Lister et al. 2009). These data are a mix of photometric and spectroscopic redshifts with varying accuracies; in any case, the accuracies are good enough for our purpose. Our AGN sample spans a wide range in redshift, from $z = 0.02$ (0316+413 = 3C84) to $z = 2.69$ (0239+108) with the peak of the distribution located at $z \simeq 0.7$. This means that our survey covers a *rest-frame* frequency range of 82–738 GHz (as far as known). We present the distribution of redshifts in Fig. 3.

Thanks to the high sensitivity of the PdBI and the fact that $q(\psi)$ is a relative parameter, our survey is highly precise. The median *statistical* errors $\delta$ in $m_L$ and $\chi$ are

$\delta m_L$ = 0.32% and $\delta\chi$ = 1.9° for the 1.3mm band,
$\delta m_L$ = 0.16% and $\delta\chi$ = 1.5° for the 2mm band,
$\delta m_L$ = 0.10% and $\delta\chi$ = 1.1° for the 3mm band.

These numbers reveal a level of precision that is unprecedented (compare, e.g., Stevens et al. 1996; Nartallo et al. 1998); they also show a tendency towards higher precision for lower frequencies. We present the error distributions in Fig. 4; for reasons of visibility, we do not separate frequency bands here but show one global distribution for $\delta m_L$ and $\delta\chi$, respectively.

### 3.2. Polarization Levels

We actually detected polarization in 316 (out of 441) measurements; the remaining 125 were non-detections resulting in polarization fraction upper limits. Out of our 86 AGN, we found for 73 of them significant polarization at least once; the remaining 13 objects remained unpolarized (within errors) during our monitoring campaign. For 45 sources we detect polarization at least twice; for seven sources, the number of positive measurements exceeds ten. This corresponds to high detection rates: for ∼72% of all measurements (∼85% of all sources) we actually see polarized emission. The polarization fractions we find range from 0.6% (1928+738 at 93.2 GHz in November 2009) to 19.4% (1005+066 at 83.3 GHz in January 2008) with a median of ∼4.1%. In Fig. 5 we present two histograms of polarization fractions: one including all measurements (top panel), one including one averaged (in time and frequency) value per source (center and bottom panel). The histograms are shown together with the best-fitting Poissonian profiles (generalized to non-integer bin values).

The distribution of source-averaged polarization fractions (Fig. 5, center panel) can roughly be described by a Poissonian distribution; a goodness-of-fit test finds $\chi^2$/dof = 1.51 (dof being the number of degrees of freedom), meaning a probability of ∼10% that the distribution is indeed intrinsically

[2] Accessible via http://nedwww.ipac.caltech.edu/
[3] See http://www.physics.purdue.edu/MOJAVE/allsources.shtml

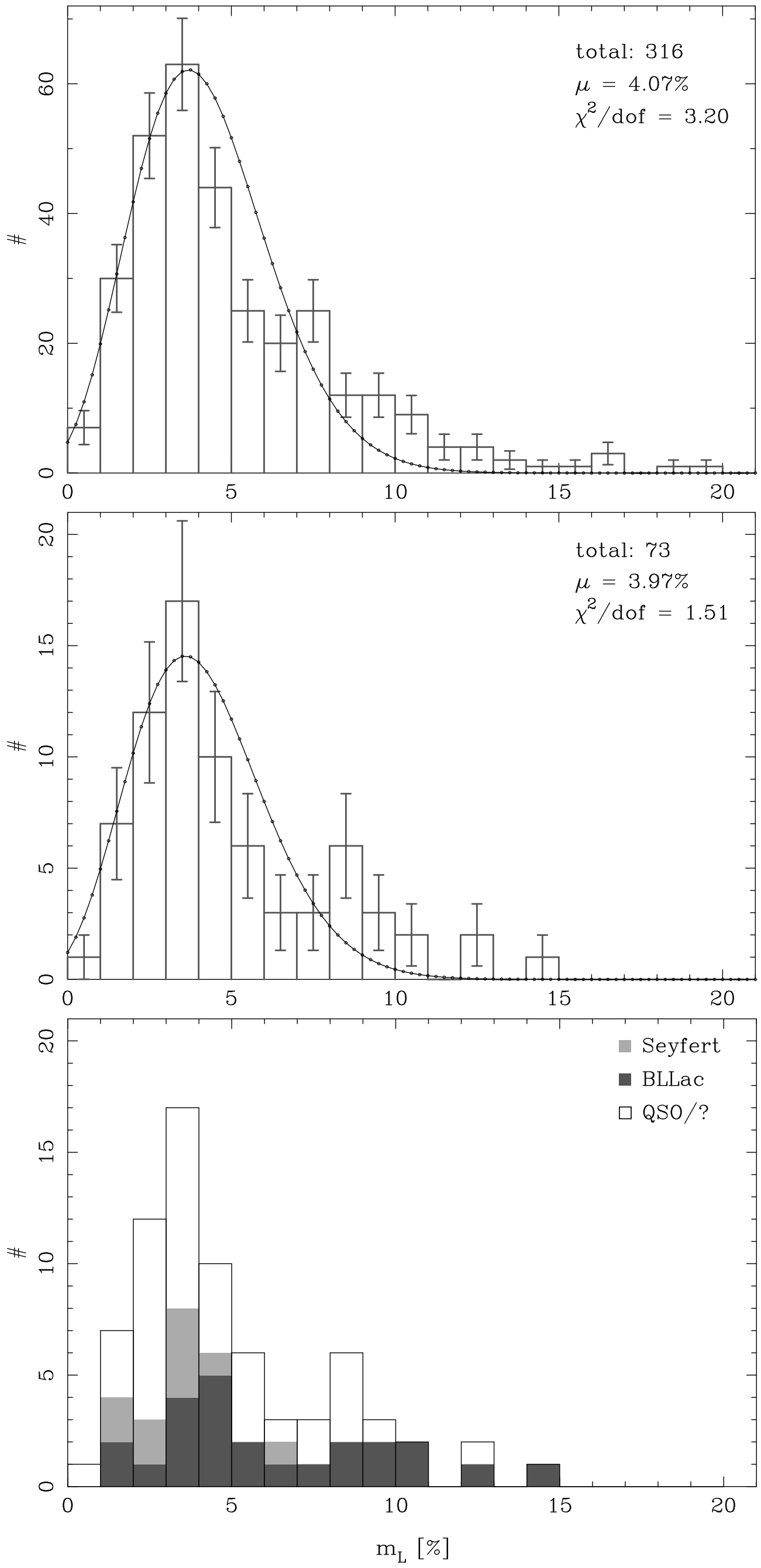

**Fig. 5.** Histograms of fractional linear polarization. Top panel: all measurements which detected linear polarization. Center panel: one averaged (over both times and frequencies) value per source. Bottom panel: same as center panel, but separating the contributions of different source types. Errorbars are binomial errors. The black curves indicate the best-fitting Poissonian profiles; the corresponding $\chi^2$/dof values are given in the plots. The median values are $\mu \simeq$ 4.1% and $\mu \simeq$ 4.0% respectively; we find polarization fractions up to ≈19%.

Poissonian. For the distribution of all measurements (Fig. 5, top panel) we find $\chi^2$/dof = 3.20, meaning a probability of agreement of less than 0.1%. This is a priori not surprising as this histogram includes multiple measurements of sources (∼4 in average), i.e. correlated data. However, in both histograms there is a marginal trend towards a pronounced tail (maybe even bimodality) of the distributions. This may indicate a systematic difference between source types.

In order to investigate this quantitatively, we grouped our sources into Seyfert galaxies, BL Lacertae (BLLac) objects, and QSOs (the last group including sources of unknown type or unclear classification). Like the redshifts, we took the source classifications from the NED and the MOJAVE archive. From all 73 AGN with detected polarization, 24 are of type BLLac, 10 are Seyfert galaxies, and 39 are QSOs/unclear. The *average* degrees of polarization are

$m_L$ = 6.5 ± 0.7% for BLLacs,

$m_L$ = 3.4 ± 0.4% for Seyferts,

$m_L$ = 4.5 ± 0.4% for QSO/unclear.

Errors are standard errors $\sigma_m/\sqrt{N}$. Although these numbers are limited by low-number statistics, they reveal a trend for BLLac objects to be the most polarized AGN in the mm wavelength range (see Fig. 5, bottom panel). Especially, there are no Seyfert galaxies with polarization fractions above 7%.

### *3.3. Polarization vs. Frequency*

As our dataset spans approximately a factor three in observed frequencies $\nu$ and a factor nine in rest-frame frequencies $\nu_0$, one needs to take into account possible correlations between polarization parameters and frequency. In Fig. 6 we present all polarization fraction measurements as function of observed (top panel) and rest-frame (bottom panel) frequency. When testing for the presence of a trend in the distribution, one needs to be cautious for two reasons. First, the dataset includes multiple measurements of sources at different frequencies, meaning there are correlations among the data points. Second, there is a bias towards higher values for $m_L$ at higher observed frequencies due to data quality: measurements at higher observed frequencies tend towards lower S/N (see Sect. 3.1), thus reducing the probability for detecting small $m_L$ values (i.e. increasing the incompleteness for small $m_L$). In the top panel of Fig. 6 this becomes visible as a "sensitivity floor" that rises linearly towards larger frequencies. For interpreting the relation between polarization and observed frequency, we therefore choose a careful approach and only take into account the ranges the data points occupy in the $\nu - m_L$ diagram. One finds that in each wavelength band the data are approximately confined to the same range of values.

Multiple measurements of the same source provide additional information. As mentioned in Sect. 3.2, we have seven sources at hand for which polarization was detected more than ten times. When inspecting the polarizations as function of frequency, it turns out that the only relevant effect is variability with time. On the one hand, degrees of polarizations tend to scatter by several per cent (i.e., by factors up to ∼2) even if measured at the same frequency. On the other hand, polarization levels measured at different frequencies are always located around the same typical value, i.e. the data are in agreement with flat lines in $\nu - m_L$ diagrams (or they just form point clouds). We will discuss the behaviour of individual sources in detail in paper II.

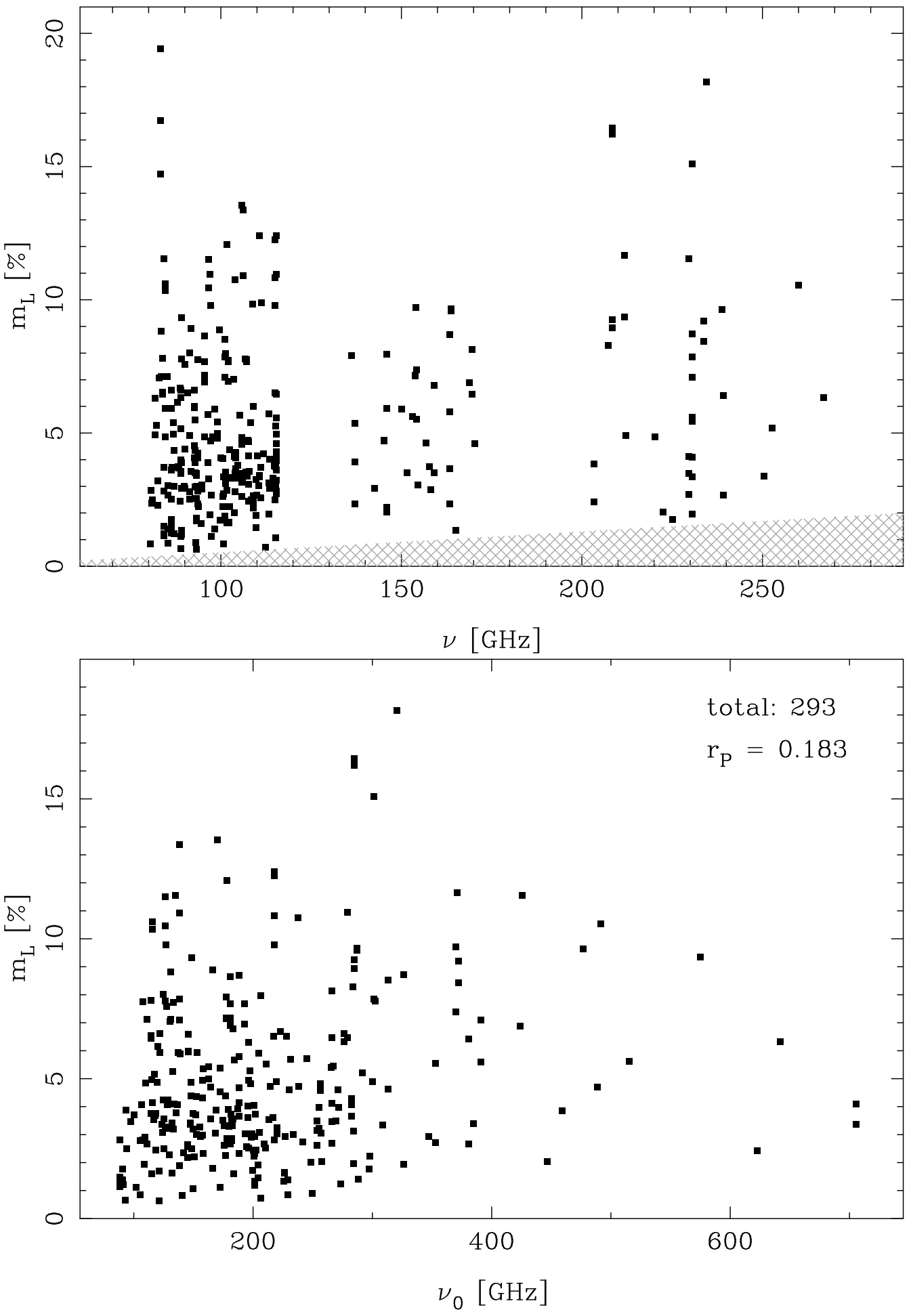

**Fig. 6.** *Top panel*: Degree of linear polarization vs. *observed* frequency. The three groups of data points correspond to the 1.3mm, 2mm, and 3mm bands of the PdBI. There is no clear correlation between frequency and polarization fraction; in each band, the data are approximately confined to the same range of values. The grey-shaded area marks the "sensitivity floor" where no data are available due to S/N constraints. *Bottom panel*: Degree of linear polarization vs. *rest-frame* frequency. This diagram includes only sorces for which a redshift is known. The Pearson coefficient $r_P = 0.183$ indicates the absence of correlation.

When inspecting the relation between polarization and *rest-frame* frequency (Fig. 6, bottom panel) there is no indication for a correlation either. We quantify the presence (or absence) of a correlation by using the Pearson correlation coefficient $r_P$ (e.g. Snedecor & Cochran 1980). For the $\nu_0 - m_L$ distribution, we find $r_P = 0.183$; this indicates the absence of an intrinsic correlation. Again, one has to take into account that this distribution contains multiple measurements of sources, meaning the data are not fully independent.

From our discussion we therefore conclude that (within our uncertainties) there is no indication for a correlation between frequency and degree of polarization in our data.

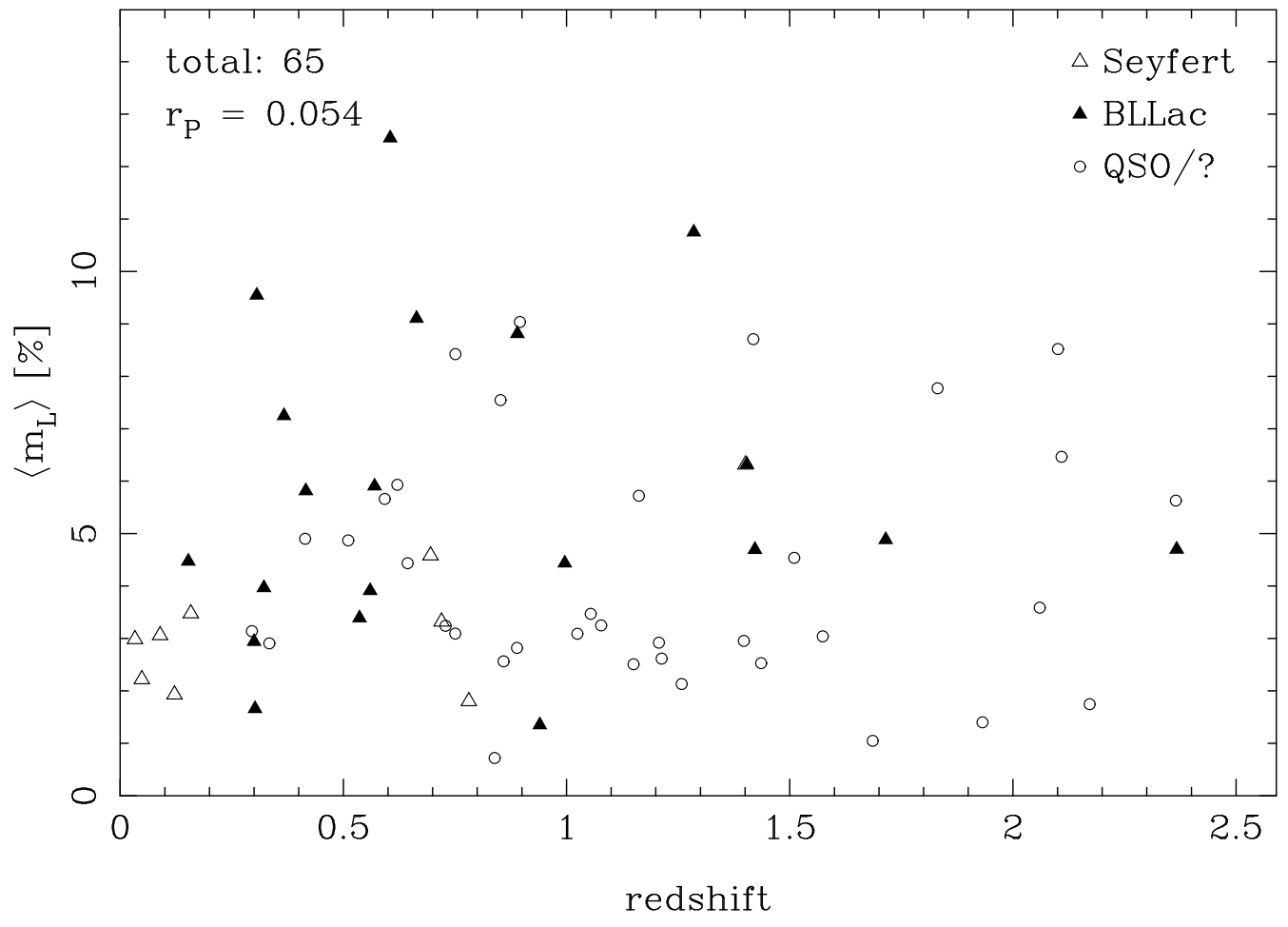

**Fig. 7.** Distribution of mean linear polarization fraction vs. source redshift. The diagram includes all sources with known redshift that have been observed to be polarized at least once; this is the case for 65 AGN. For each source one polarization value averaged over time and frequency is shown. According to their classification, sources are labeled as Seyfert galaxy, BLLac object, or QSO/unclear ("QSO/?"). The Pearson coefficient $r_P = 0.054$ indicates the absence of a correlation between redshift and polarization level. One should note that many sources show variability in $m_L$ of several per cent; we do not indicate those ranges in the diagram for reasons of visibility.

### 3.4. Polarization vs. Redshift

For 65 out of our 86 AGN, we have detected polarization at least once *and* have redshift information at hand. In Fig. 7 we present the distribution of source-averaged polarization fractions $\langle m_L \rangle$ vs. source redshifts $z$. We additionally grouped our sources into Seyfert galaxies, BLLac objects, and QSOs/unclear. From visual inspection of the $\langle m_L \rangle - z$ diagram, there appears to be no obvious patter in the distribution. For the distribution shown in Fig. 7, we find a Pearson correlation coefficient $r_P = 0.054$, clearly indicating the absence of a correlation between redshift and degree of polarization. We note that the maximum redshift of Seyfert galaxies in our sample is $z = 1.4$, well below the maximum redshifts of the other source types.

### 3.5. Polarization vs. Spectral Index

Combining our polarization data with flux information can provide valuable physical information. For this study we make use of flux data from the PdBI calibrator flux monitoring program (Krips et al. *in prep.*). As for many sources the PdBI flux monitoring provides frequency-resolved flux density information, we computed spectral indices where possible. For this we made use of the standard assumption that the flux density $S_\nu$ is related to the frequency $\nu$ via a power law $S_\nu \propto \nu^{-\alpha}$ with $\alpha$ being the spectral index.

Our analysis is complicated by two effects: (a) it is not possible to observe several frequencies simultaneously with the PdBI, meaning we have to use non-simultaneous flux data; (b) our targets show strong intrinsic variabilities up to ~60% (fractional standard deviation) at any given frequency. These effects mean that the $\nu - S_\nu$ relations we analyze show a lot of scatter which prevents us from testing models more complicated than simple power laws (like, e.g., spectral turnovers, or broken power laws). Within these limitations, we have not found a case for which the simple power law model is not sufficient for describing the data.

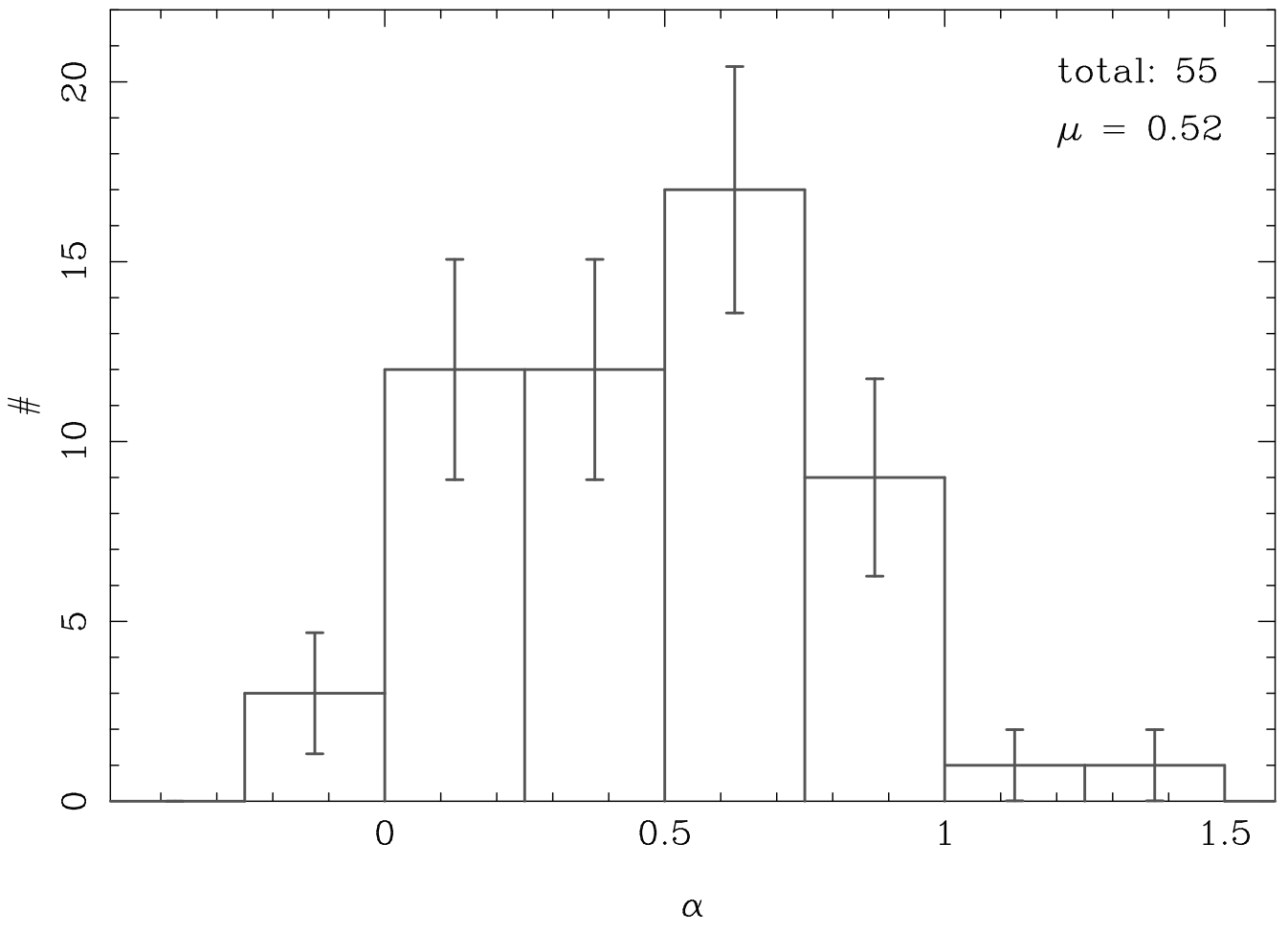

**Fig. 8.** Histogram of spectral indices $\alpha$ for 55 of our AGN. The index is defined via $S_\nu \propto \nu^{-\alpha}$. The distribution ranges from -0.2 to +1.3 with a median value $\mu = 0.52$. Errorbars indicate binomial errors.

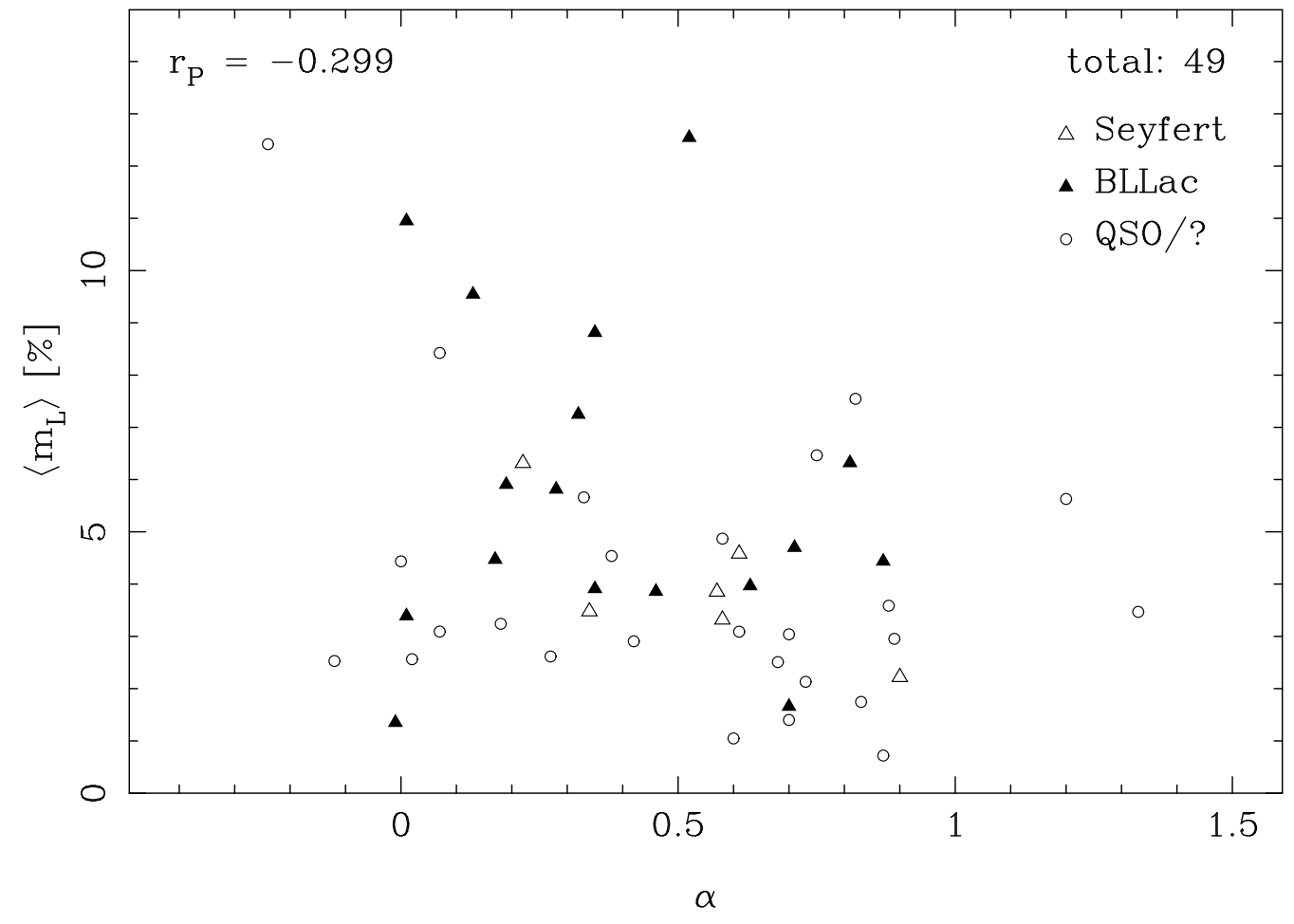

**Fig. 9.** Source-averaged degree of linear polarization $\langle m_L \rangle$ vs. spectral index $\alpha$ (defined via $S_\nu \propto \nu^{-\alpha}$). Here we show those 49 sources for which a value for $\alpha$ is available and which have been found to be polarized. The correlation coefficient $r_P = -0.299$ indicates that there is no strict correlation between spectral index and degree of polarization, but that there is a trend towards sources with lower values of $\alpha$ having higher polarization levels. One should note that (a) many sources show variability in $m_L$ of several per cent, and (b) the median error in $\alpha$ is 0.19. We do not indicate those ranges in the diagram for reasons of visibility.

We computed $\alpha$ by means of a least-squares fit. We performed the calculation for those 55 sources for which at least seven measurements (because the measurements are non-simultaneous) covering a frequency range of 50 GHz or more (in order to ensure a reasonable accuracy of the slope) are available. The substantial scatter in the $S_\nu$ data leads to quite large formal statistical errors $\delta\alpha$. We find a median uncertainty of 0.19, with the largest value being $\delta\alpha = 0.82$.

In Fig. 8 we present a histogram of the spectral indices. The distribution covers a range in $\alpha$ from -0.2 to +1.3. The median spectral index is located at $\alpha = 0.52$. This shows that our sample is composed of sources with spectra that are preferentially flat or of moderate steepness. In order to test a potential relation between spectral index and polarization properties, we inspected the source-averaged polarization fraction $\langle m_L \rangle$ as function of $\alpha$. The corresponding diagram is presented in Fig. 9. We find a Pearson correlation coeficient $r_P = -0.299$; this indicates that there is no strict correlation between spectral index and polarization fraction. However, this value points out a trend towards sources with lower $\alpha$ having larger polarization fractions. This is actually visible by eye: for $\alpha < 0.5$, the data span a range in polarization fractions up to ~12.5%; for $\alpha > 0.5$, there are no data beyond $\langle m_L \rangle = 7.5\%$. We do not see a relation between source class (Seyfert, BLLac, QSO) and location in the diagram (except of the – already discussed – fact that BLLac objects tend to be those with the highest polarization).

We summarize our survey in Table 1. For each source we give B1950 name, redshift, spectral indices and their uncertainties, fluxes and flux variabilities, polarization fractions and their variabilities, and polarization angles and their variabilities as far as available. According to their frequencies, measurements are grouped into 1.3mm, 2mm, and 3mm wavelength bands. Please note that we quote source variabilities $\sigma_X$; these values are standard deviations over measurements from different dates. These are *not* accuracies of individual measurements, which are usually much smaller than the intrinsic variabilities (cf. Sect. 3.1 and Fig. 4). Entries "–" denote "not observed". If a source was observed but no polarization could be detected, we quote the 3-$\sigma$ upper limit on $m_L$; those values are denoted by "<".

## 4. Discussion

Our survey provides an overview over the polarimetric properties of AGN at millimeter wavelengths. In order to put our results into context, one needs to take into account that our source sample is biased with respect to brightness. As our targets serve as calibrators, there is a preferential selection of bright AGN located close to primary PdBI targets. Our sources are mm/radio-bright ($S_\nu \gtrsim 0.2$ Jy at 3mm; see Table 1) and have (as far as known) spectra that are consistent with power law spectra. We therefore assume in the following that the emission we observe is fully non-thermal.

For synchrotron emission from an ensemble of homogeneous and isotropic relativistic electrons moving in a uniform magnetic field (B-field), the degree of linear polarization can be calculated analytically. If the energy distribution of the electrons follows a power law with index $\Gamma$, then $\Gamma = 2\alpha + 1$ and

$$m_L = \frac{\Gamma + 1}{\Gamma + 7/3} \tag{5}$$

(e.g. Pacholczyk 1970) for optically thin sources. For flat-spectrum sources with $\alpha \sim 0$ (i.e. $\Gamma \sim 1$), one finds degrees of polarizations $m_L \sim 60\%$; if $\alpha > 0$, $m_L$ increases even further. Obviously, our observations find much smaller values (see e.g. Fig. 5); even our highest value, $m_L \simeq 19\%$, is more than a factor of three below the theoretical number. Clearly, the AGN we observe are far away from being "ideal" synchrotron emitters at mm wavelengths. Instead, the superposition of emission from various components (cores, lobes, jets) not resolved by the PdBI (within its present range of baselines) seems to average out most of the polarization signal. For a given component, e.g. a jet, averaging due to non-uniform B-field geometries can be expected to reduce the observed polarization signal further. This is in good agreement with earlier results (e.g. Nartallo et al. 1998). An additional explanation may be that some emission regions are actually optically thick. In this case,

$$m_L = \frac{1}{2\Gamma + 13/3} \qquad (6)$$

(e.g. Pacholczyk 1970). For $\alpha \sim 0$, one finds $m_L \sim 16\%$. If $\alpha$ increases, the polarization level *decreases* further, resulting in values located in the $m_L$ range we actually observe.

The distribution of the source-averaged polarization fractions (see Fig. 5) provides two conclusions. First, we do see a trend that BLLac sources have higher intrinsic polarization levels. Second, there is no obvious segregation of source types; the distributions show a large amount of overlap, indicating a smooth transition between source types.

The absence of a correlation between redshift and (source-averaged) degree of polarization (see Fig. 7) provides interesting constraints on the time evolution of AGN. As we observe polarized emission in 65 sources with $z \lesssim 2.4$, our sample probes a substantial part of the age of the universe. From this we can conclude that the properties of polarized mm radiation from AGN have not evolved substantially since $z \simeq 2.4$.

The spectral indices of our sources are located in the range $\alpha \simeq -0.2...1.3$ (see Fig. 8). This indicates that our sample actually provides a mix of core-dominated ($\alpha \lesssim 0.5$) and outflow (lobe/jet)-dominated ($\alpha \sim 0.5-1$; e.g. Krolik 1999) sources. On the one hand, the absence of a clear correlation between spectral index and degree of polarization (see Fig. 9) suggests that (at least at mm wavelengths) the polarization levels of AGN cores and outflows are of the same order of magnitude. On the other hand, there is a marginal trend towards higher polarization in core-dominated sources; this leaves room for systematically higher polarization levels in AGN cores compared to outflows by factors $\approx$2.

Our results are in good agreement with AGN properties observed in the centimeter wavelength range, with respect to the distributions of polarization fractions and spectral indices obtained from various samples. The surveys by Altschuler & Wardle (1976, 1977) and Aller et al. (1985) found maximum polarizations of ~15% in the frequency range 2–14.5 GHz, with "typical" degrees of polarization being $\lesssim$5%. This is especially interesting as our survey probes the high-frequency end of the "pure synchrotron" domain of the AGN spectral energy distribution. At frequencies above ~1 THz, other radiation processes like Compton scattering in the accretion zone around the central black hole can contribute substantially to the emission (e.g. Krolik 1999 and references therein). Observationally, this transition expresses itself in the "millimeter-break" located at $\nu \gtrsim 1$ THz (e.g. Elvis et al. 1994). We find for our frequency range the same characteristic radiation properties – in terms of polarization and distribution of spectral indices – as for the cm wavelength range. From this we can conclude that the rest-frame frequency range up to 738 GHz traces – like the cm/radio continuum domain – the signatures of "pure" synchrotron emission without noticeable contributions from other emission mechanisms. The mm-break appears to be not detectable in the frequency range covered by our survey. A one-to-one comparison of results for given sources across various studies is beyond the scope of this article. However, we are going to discuss individual sources and corresponding multi-wavelegth observations in detail in paper II.

We find that substantial variations in the degree of polarization (up to ~60% in fractional standard deviation) on timescales from weeks to years are a common feature throughout our AGN sample (see Table 1). This is the case also for the source fluxes that show variabilities at a similar level.

Any discussion of variability in polarization is clearly limited by the sparse and irregular sampling (both in time and frequency) of our polarization data (see also Sect. 3.2). However, in flux, variability on various time scales is a well-known feature and has been explored in long-term studies, notably the works by Hovatta et al. (2007, 2008) that cover time lines of 25 years. From a detailed statistical analysis they conclude that (a) characteristic time scales for variations are ~1 to 6 years depending on the strengths of the variations, (b) variations are the more seldom the stronger they are, (c) there is a characteristic flare length of ~2.5 years, and (d) source variability and characteristic time scales are not related to source type (QSO, BLLac, galaxy).

From this we can conclude that (1) the variability we observe in polarization levels fits the present picture of AGN variability, but that (2) our survey probably has not yet reached the time coverage to be sensitive to the characteristic time scales of variations which have been seen in light curves. It is interesting to note that Hovatta et al. (2007, 2008) find no relation between flux variability and source type, whereas we find a relation between polarization *level* and source type (see Sect. 3.2). It will therefore be important to test for the presence of a relation between polarization *variability* and source type as soon as this becomes feasible. We are going to discuss this in paper II which deals with sources for which we have multiple (>10) polarization measurements at hand.

High variabilities in both, fluxes and polarization fractions, point towards strong variations in source activities, accretion flows, B-field geometries, and/or outflows (jets) on the timescales of our survey. The most common explanation of such violent behaviour are shocks occuring in relativistic jets, possibly accompanied by jet bending and changes in B-field configurations (e.g. Stevens et al. 1998; Nartallo et al. 1998). Other, more recent models propose the orbital motion of plasma concentrations ("hotspots") in the accretion disks of supermassive black holes in order to explain flux and polarization variations on timescales from hours to weeks (e.g. Broderick & Loeb 2006). In any case, our results indicate that these processes are a frequent phenomenon for luminous AGN.

## 5. Summary and Conclusions

We have made use of IRAM/PdBI calibration measurements in order to collect polarimetric data for 86 active galactic nuclei in the *observed* frequency range 80–267 GHz, corresponding to a *rest-frame* frequency range 82–738 GHz. We obtain 441 measurements in total and find significant linear polarization in 316 measurements belonging to 73 sources. Our survey achieves unprecedented precisions with median errors of ~0.1% in polarization fractions and ~1.2° in polarization angles. Additionally, we collect source flux data from the PdBI calibrator monitoring. The main results are:

- The median polarization fraction for our sources is ~4%, the highest value found is ~19%. These numbers are far below the value $\gtrsim$60% expected from "ideal" synchrotron emission from optically thin sources, pointing towards a substantial averaging of polarization due to limited spatial resolution and non-uniform B-field geometries, and maybe also emission from optically thick source regions.
- The source-averaged degrees of polarization follow approximately a Poisson distribution with a slight trend towards a pronounced tail or bimodality. In average, BLLac sources show a larger degree of polarization ($m_L \simeq 7\%$) than Seyfert galaxies or QSOs ($m_L \simeq 3\%$ and $m_L \simeq 5\%$, respectively).
- The spectral indices of our AGN are located in the range $\alpha \sim -0.2$ to $+1.3$, indicating that our sample is a mix of core- and outflow-dominated sources. There is no clear correlation between polarization and spectral index, but there is a trend towards higher polarization fractions in core-dominated sources.
- We do not find correlations of polarization with rest-frame frequencies (shown by a Pearson correlation coefficient $r_P = 0.183$) or redshifts. The absence of a $m_L - z$ correlation ($r_P = 0.054$) indicates that there has been no noticable change of polarization properties since $z \simeq 2.4$.
- The values and distributions we find for $m_L$ and $\alpha$ are in good agreement with results obtained from various samples and studies in the centimeter wavelength range (though we do attempt a one-to-one comparison of sources here). This indicates that the rest-frame frequency range we probe ($\nu_0 \leq 738$ GHz) can be described as a continuation of the cm/radio continuum domain for which synchrotron radiation is the only noteworthy emission mechanism. Especially, the mm-break appears to be not detectable in our frequency range.
- The high variabilities of polarization fractions and source fluxes (up to ~60%) strengthen the picture that violent processes – like shocks in relativistic jets or plasma "hotspots" moving in accretion disks – are a frequent phenomenon.

Our analysis provides a valuable overview on the properties of luminous AGN in the millimeter radio range. The next step will be the detailed analysis of selected individual sources in the forthcoming paper II. Our data contain multiple measurements of sources, in seven cases the number of measurements exceeds ten. These numbers will grow as our survey continues with ongoing operation of the PdBI. Our analysis is a symbiotic "backpack" project on top of the standard PdBI operations; therefore, we do not need to allocate dedicated observing time but can (in principle) continue our survey as long as we wish to. Presently, we collect ~170 polarization measurements per year, meaning that our data base is going to grow rapidly. We therefore expect valuable new insights in the near future.

*Acknowledgements.* We would like to thank Jan Martin Winters (IRAM Grenoble) for his help with the treatment of PdBI data. We are grateful to Ivan Agudo (IAA Granada) and Thomas Krichbaum (MPIfR Bonn) for helpful discussions on polarimetry and AGN polarization.

Our work has made use of the NASA/IPAC Extragalactic Database (NED) which is operated by the Jet Propulsion Laboratory, California Institute of Technology, under contract with the National Aeronautics and Space Administration.

We have made use of data from the MOJAVE database that is maintained by the MOJAVE team (Lister et al. 2009).

Last but not least, we are grateful to the reviewer, E. Ros, who helped to improve the quality of this article.

**Table 1.** Fluxes and polarization parameters for our sample of 80 AGN. Data are grouped into 1.3mm, 2mm, and 3mm bands according to their frequencies; data belonging to a given band are averaged in time and frequency. Source types are Seyfert galaxies (Sy), High (Low) Polarization Quasars (HPQ/LPQ), BL Lacertae objects (BLLac), and Flat Spectrum Radio Quasars (FSRQ) as taken from the NED and MOJAVE data bases. The parameter $z$ is the redshift, $\alpha$ the spectral index (defined via $S_\nu \propto \nu^{-\alpha}$), $S_\nu$ the flux density, $m_L$ the fraction of linear polarization, and $\chi$ is the polarization angle. The $\sigma_X$ are the source variabilities (standard deviation over all measurements from different dates) of parameters $X$; these are not measurement errors, which are usually much smaller than the intrinsic scatter of the data. If only one measurement is available, we quote the statistical error of $X$ as $\sigma_X$ instead. Signs "–" denote "not observed". Numbers with "<" are upper limits, to be read as "observed, but no polarization detected".

| | | | | | 1.3mm band | | | | | | 2mm band | | | | | | 3mm band | | | | | |
|---|---|---|---|---|---|---|---|---|---|---|---|---|---|---|---|---|---|---|---|---|---|---|
| name (B1950) | type | $z$ [1] | $\alpha$ [1] | $\sigma_\alpha$ [1] | $S_\nu$ [Jy] | $\sigma_S$ [Jy] | $m_L$ [%] | $\sigma_m$ [%] | $\chi$ [°] | $\sigma_\chi$ [°] | $S_\nu$ [Jy] | $\sigma_S$ [Jy] | $m_L$ [%] | $\sigma_m$ [%] | $\chi$ [°] | $\sigma_\chi$ [°] | $S_\nu$ [Jy] | $\sigma_S$ [Jy] | $m_L$ [%] | $\sigma_m$ [%] | $\chi$ [°] | $\sigma_\chi$ [°] |
| 0007+106 | Sy 1.2 | 0.09 | – | – | – | – | – | – | – | – | – | – | 3.1 | 0.1 | 95 | 1 | 1.6 | 1.0 | <1.5 | – | – | – |
| 0007+171 | FSRQ | 1.60 | – | – | – | – | – | – | – | – | – | – | – | – | – | – | 0.3 | 0.1 | <4.9 | – | – | – |
| 0059+581 | QSO | 0.64 | 0.0 | 0.2 | 1.9 | 1.0 | – | – | – | – | 2.4 | 0.5 | 4.7 | 0.2 | 72 | 1 | 2.1 | 0.9 | 4.4 | 1.2 | 65 | 46 |
| 0106+013 | HPQ | 2.10 | – | – | – | – | – | – | – | – | – | – | – | – | – | – | 0.7 | 0.1 | <3.3 | – | – | – |
| 0119+115 | BLLac | 0.57 | 0.2 | 0.1 | 1.4 | 0.4 | – | – | – | – | 1.2 | 0.2 | 7.3 | 1.2 | 12 | 6 | 1.5 | 0.3 | 5.6 | 2.2 | 123 | 55 |
| 0122−003 | LPQ | 1.08 | – | – | – | – | – | – | – | – | – | – | – | – | – | – | 0.4 | 0.1 | 3.3 | 0.2 | 57 | 4 |
| 0212+735 | BLLac | 2.37 | 0.7 | 0.1 | 0.9 | 0.2 | – | – | – | – | 1.1 | 0.2 | 4.7 | 0.3 | 144 | 2 | 1.5 | 0.6 | – | – | – | – |
| 0215+015 | BLLac | 1.72 | – | – | 0.8 | 0.2 | 9.4 | 1.1 | 18 | 2 | – | – | – | – | – | – | 1.0 | 0.1 | 3.8 | 0.5 | 172 | 3 |
| 0221+067 | HPQ | 0.51 | 0.6 | 0.5 | 0.4 | 0.4 | – | – | – | – | 0.4 | 0.1 | – | – | – | – | 0.5 | 0.2 | 4.9 | 0.3 | 93 | 1 |
| 0234+285 | QSO | 1.21 | 0.3 | 0.3 | 1.5 | 0.3 | – | – | – | – | – | – | – | – | – | – | 2.1 | 0.7 | 2.6 | 1.2 | 140 | 12 |
| 0235+164 | BLLac | 0.94 | 0.0 | 0.3 | 2.1 | 0.9 | – | – | – | – | 1.8 | 0.1 | – | – | – | – | 2.1 | 1.2 | 1.4 | 0.3 | 176 | 34 |
| 0239+108 | QSO | 2.69 | 0.7 | 0.4 | 0.4 | 0.2 | – | – | – | – | 0.5 | 0.1 | <4 | – | – | – | 0.5 | 0.1 | – | – | – | – |
| 0316+413 | Sy 2 | 0.02 | 0.6 | 0.1 | 4.8 | 1.6 | <1.8 | – | – | – | 6.3 | 1.3 | – | – | – | – | 7.7 | 1.7 | <0.7 | – | – | – |
| 0333+321 | LPQ | 1.26 | 0.7 | 0.2 | 0.7 | 0.3 | 3.9 | 0.4 | 164 | 1 | – | – | – | – | – | – | 1.3 | 0.5 | 1.3 | 0.1 | 135 | 2 |
| 0336−019 | HPQ | 0.85 | 0.8 | 0.8 | 0.6 | 0.4 | 11.5 | 1.0 | 103 | 1 | 1.1 | 0.1 | <3.4 | – | – | – | 0.6 | 0.1 | 3.6 | 0.1 | 102 | 1 |
| 0355+508 | LPQ | 1.51 | 0.4 | 0.2 | 4.2 | 1.5 | – | – | – | – | 5.5 | 0.9 | 6.9 | 0.1 | 5 | 1 | 6.4 | 1.8 | 4.1 | 2.0 | 177 | 9 |
| 0400+258 | FSRQ | 2.11 | 0.8 | 0.2 | 0.4 | 0.1 | – | – | – | – | 0.3 | 0.1 | – | – | – | – | 0.8 | 0.1 | 6.5 | 0.2 | 99 | 1 |
| 0415+379 | Sy 1 | 0.05 | 0.9 | 0.1 | 4.5 | 2.8 | <3.3 | – | – | – | 3.5 | 1.5 | 5.9 | 0.1 | 106 | 1 | 8.8 | 3.8 | 1.9 | 1.2 | 164 | 40 |
| 0420−014 | BLLac | 0.91 | 0.5 | 0.1 | 2.2 | 0.7 | – | – | – | – | 2.9 | 0.7 | – | – | – | – | 3.5 | 0.8 | <3.0 | – | – | – |
| 0430+052 | Sy 1 | 0.03 | – | – | – | – | – | – | – | – | 1.4 | 0.2 | 2.2 | 0.3 | 167 | 2 | 1.8 | 0.1 | 3.8 | 0.2 | 170 | 2 |
| 0444+634 | Sy 1 | 0.78 | – | – | – | – | – | – | – | – | – | – | – | – | – | – | 0.4 | 0.1 | 1.8 | 0.1 | 130 | 2 |
| 0446+112 | QSO? | 1.21 | – | – | – | – | – | – | – | – | – | – | – | – | – | – | 0.6 | 0.2 | 2.9 | 0.1 | 154 | 1 |
| 0507+179 | BLLac? | 0.42 | 0.3 | 0.1 | 0.8 | 0.3 | 5.3 | 4.8 | 72 | 34 | 1.6 | 0.1 | 8.0 | 0.1 | 81 | 1 | 1.0 | 0.3 | 5.7 | 1.7 | 110 | 14 |
| 0528+134 | LPQ | 2.06 | 0.9 | 0.1 | 2 | 1.1 | 3.3 | 0.8 | 67 | 34 | 1.7 | 0.5 | 2.0 | 0.1 | 18 | 2 | 3.6 | 1.3 | 4.0 | 1.2 | 75 | 56 |
| 0529+483 | FSRQ | 1.16 | – | – | – | – | – | – | – | – | 0.9 | 0.1 | – | – | – | – | 1.0 | 0.2 | 5.7 | 0.1 | 122 | 1 |
| 0539−057 | HPQ | 0.84 | 0.9 | 0.3 | 0.4 | 0.2 | – | – | – | – | – | – | – | – | – | – | 0.8 | 0.2 | 0.7 | 0.1 | 176 | 3 |
| 0552+398 | LPQ | 2.37 | 1.2 | 0.2 | 0.4 | 0.1 | – | – | – | – | 0.7 | 0.1 | 5.6 | 0.4 | 140 | 2 | 1.0 | 0.1 | <2.0 | – | – | – |
| 0611+131 | QSO | 0.75 | – | – | – | – | – | – | – | – | – | – | – | – | – | – | 0.2 | 0.1 | <4.9 | – | – | – |
| 0716+714 | BLLac | 0.30 | – | – | – | – | – | – | – | – | – | – | – | – | – | – | 1.3 | 0.7 | 2.9 | 1.3 | 78 | 33 |
| 0727−115 | QSO | 1.59 | – | – | – | – | – | – | – | – | – | – | – | – | – | – | 3.8 | 0.3 | <2.7 | – | – | – |
| 0748+126 | LPQ | 0.89 | – | – | 2.3 | 0.5 | – | – | – | – | – | – | – | – | – | – | – | – | 2.8 | 0.1 | 40 | 2 |
| 0804+499 | HPQ | 1.44 | −0.1 | 0.3 | 0.6 | 0.1 | – | – | – | – | – | – | – | – | – | – | 0.4 | 0.1 | 2.5 | 0.1 | 45 | 2 |
| 0820+560 | FSRQ | 1.42 | – | – | – | – | – | – | – | – | – | – | – | – | – | – | 0.5 | 0.1 | 8.7 | 3.2 | 172 | 10 |
| 0833+585 | QSO | 2.10 | – | – | – | – | – | – | – | – | – | – | – | – | – | – | 0.3 | 0.1 | 8.5 | 0.2 | 152 | 1 |
| 0836+710 | LPQ | 2.17 | 0.8 | 0.4 | 0.4 | 0.1 | – | – | – | – | 1.7 | 0.1 | – | – | – | – | 1.1 | 0.2 | 1.7 | 0.7 | 118 | 19 |
| 0851+202 | BLLac | 0.31 | 0.1 | 0.2 | 4.0 | 1.8 | 11.5 | 5.1 | 173 | 1 | 5.3 | 1.4 | 7.9 | 0.3 | 147 | 5 | 4.5 | 1.7 | 9.3 | 3.1 | 163 | 5 |
| 0906+015 | HPQ | 1.02 | 0.6 | 0.2 | 0.5 | 0.1 | – | – | – | – | 0.8 | 0.1 | <4.7 | – | – | – | 0.9 | 0.2 | 3.1 | 0.1 | 121 | 3 |
| 0923+392 | Sy 1 | 0.70 | 0.6 | 0.1 | 2.6 | 0.6 | 6.3 | 1.1 | 159 | 1 | 3.4 | 0.6 | 3.5 | 0.1 | 150 | 2 | 4.4 | 0.6 | 4.0 | 0.5 | 151 | 3 |
| 0954+556 | HPQ | 0.90 | – | – | – | – | – | – | – | – | – | – | – | – | – | – | 0.5 | 0.1 | 9.0 | 2.0 | 8 | 2 |
| 0954+658 | BLLac | 0.37 | 0.3 | 0.1 | 0.8 | 0.3 | 12.9 | 4.5 | 177 | 4 | 1.4 | 0.4 | 4.2 | 2.1 | 11 | 17 | 1.2 | 0.4 | 6.0 | 2.4 | 167 | 38 |
| 1005+066 | BLLac | – | – | – | – | – | – | – | – | – | – | – | – | – | – | – | 0.4 | 0.1 | 14.7 | 4.9 | 152 | 4 |
| 1012+232 | Sy 1.5 | 0.56 | – | – | – | – | – | – | – | – | – | – | – | – | – | – | 0.6 | 0.1 | <2.3 | – | – | – |
| 1030+611 | Sy 1 | 1.40 | 0.2 | 0.2 | 0.3 | 0.1 | – | – | – | – | 0.3 | 0.1 | 8.5 | 1.6 | 163 | 2 | 0.3 | 0.2 | 5.4 | 0.6 | 95 | 31 |
| 1040+244 | BLLac | 0.56 | 0.4 | 0.2 | 0.7 | 0.2 | – | – | – | – | – | – | – | – | – | – | 0.9 | 0.2 | 3.9 | 0.5 | 125 | 4 |
| 1044+719 | FSRQ | 1.15 | 0.7 | 0.2 | 0.4 | 0.1 | – | – | – | – | 0.6 | 0.1 | <5.1 | – | – | – | 0.7 | 0.2 | 2.5 | 0.2 | 6 | 11 |
| 1055+018 | BLLac | 0.89 | 0.4 | 0.1 | 2.5 | 0.5 | 10.5 | 0.3 | 129 | 3 | 2.7 | 0.3 | – | – | – | – | 3.3 | 0.7 | 8.4 | 2.7 | 126 | 4 |
| 1144+402 | QSO | 1.09 | – | – | – | – | – | – | – | – | – | – | – | – | – | – | 0.7 | 0.2 | <4.4 | – | – | – |
| 1150+497 | FSRQ | 0.33 | 0.4 | 0.1 | 0.5 | 0.1 | – | – | – | – | 0.6 | 0.1 | – | – | – | – | 0.7 | 0.1 | 2.9 | 0.8 | 49 | 31 |
| 1156+295 | HPQ | 0.73 | 0.2 | 0.3 | 0.7 | 0.1 | – | – | – | – | – | – | – | – | – | – | 1.0 | 0.6 | 3.2 | 0.9 | 178 | 39 |
| 1226+023 | Sy 1 | 0.16 | 0.3 | 0.1 | 12.4 | 3.5 | 3.6 | 1.3 | 158 | 10 | 15 | 4.3 | – | – | – | – | 16.5 | 4.6 | 3.4 | 0.7 | 140 | 8 |
| 1253−055 | BLLac | 0.54 | 0.0 | 0.2 | 9.5 | 1.5 | 3.4 | 0.3 | 63 | 4 | 8.6 | 1.3 | – | – | – | – | 10.0 | 2.7 | – | – | – | – |
| 1300+580 | ? | – | −0.2 | 0.6 | 0.2 | 0.1 | – | – | – | – | 0.2 | 0.1 | – | – | – | – | 0.2 | 0.1 | 12.4 | 0.5 | 167 | 1 |
| 1307+121 | BLLac | – | 0.0 | 0.6 | 0.2 | 0.1 | – | – | – | – | – | – | – | – | – | – | 0.2 | 0.1 | 11.0 | 0.1 | 32 | 1 |
| 1308+326 | BLLac | 1.00 | 0.9 | 0.2 | 1.2 | 0.8 | 9.6 | 0.3 | 159 | 4 | 1.4 | 0.1 | 4.6 | 0.1 | 33 | 1 | 2.1 | 0.3 | 3.7 | 1.2 | 32 | 14 |
| 1334−127 | BLLac | 0.54 | 0.7 | 0.5 | 3.8 | 0.8 | <5.1 | – | – | – | – | – | – | – | – | – | 5.6 | 2.3 | – | – | – | – |
| 1345+125 | Sy 2 | 0.12 | – | – | – | – | – | – | – | – | – | – | – | – | – | – | 0.3 | 0.1 | 1.9 | 0.1 | 146 | 5 |
| 1354+195 | Sy 1.5 | 0.72 | 0.6 | 0.2 | 0.5 | 0.1 | – | – | – | – | – | – | – | – | – | – | 0.9 | 0.2 | 3.3 | 0.2 | 80 | 2 |
| 1413+135 | BLLac | 0.25 | – | – | 0.6 | 0.1 | – | – | – | – | – | – | – | – | – | – | 1 | 0.1 | <0.9 | – | – | – |
| 1417+385 | FSRQ | 1.83 | – | – | – | – | – | – | – | – | – | – | – | – | – | – | 0.2 | 0.1 | 7.8 | 0.4 | 116 | 3 |
| 1418+546 | BLLac | 0.15 | 0.2 | 0.1 | 0.8 | 0.3 | 5.2 | 0.4 | 127 | 7 | 0.6 | 0.2 | 4.9 | 2.2 | 121 | 14 | 0.7 | 0.2 | 3.9 | 1.5 | 115 | 13 |
| 1538+149 | BLLac | 0.61 | 0.5 | 0.3 | – | – | – | – | – | – | 0.4 | 0.1 | – | – | – | – | 0.7 | 0.1 | 12.5 | 1.4 | 144 | 4 |
| 1546+027 | HPQ | 0.41 | – | – | – | – | 4.9 | 0.3 | 29 | 4 | – | – | – | – | – | – | 1.4 | 0.1 | – | – | – | – |
| 1548+056 | BLLac | 1.42 | – | – | – | – | – | – | – | – | – | – | – | – | – | – | 1.5 | 0.2 | 4.7 | 0.2 | 157 | 1 |
| 1611+343 | LPQ | 1.40 | 0.9 | 0.2 | 0.5 | 0.1 | – | – | – | – | – | – | – | – | – | – | 1.1 | 0.2 | 3.0 | 0.6 | 162 | 55 |
| 1633+382 | LPQ | 1.81 | 0.3 | 0.2 | 1.8 | 0.7 | <4.9 | – | – | – | 2.1 | 0.3 | – | – | – | – | 2.5 | 0.6 | – | – | – | – |
| 1637+574 | LPQ | 0.75 | 0.1 | 0.3 | 1.4 | 0.1 | – | – | – | – | 0.7 | 0.1 | – | – | – | – | 1.3 | 0.4 | 3.1 | 0.9 | 117 | 30 |
| 1641+399 | HPQ | 0.59 | 0.3 | 0.1 | 2.8 | 1.1 | 6.7 | 2.9 | 93 | 55 | 3.7 | 1.2 | 3.8 | 1.2 | 28 | 36 | 4.0 | 1.4 | 5.4 | 0.2 | 77 | 1 |
| 1642+690 | HPQ | 0.75 | 0.1 | 0.1 | 2.3 | 0.5 | 11.7 | 0.1 | 136 | 1 | 2.9 | 0.4 | 9.6 | 0.1 | 135 | 2 | 2.9 | 0.8 | 7.0 | 3.7 | 155 | 27 |
| 1655+077 | QSO | 0.62 | – | – | – | – | – | – | – | – | – | – | – | – | – | – | 1.2 | 0.1 | 5.9 | 0.1 | 130 | 1 |
| 1730−130 | BLLac | 0.90 | 0.0 | 0.4 | 2.4 | 0.2 | – | – | – | – | 3.3 | 0.6 | – | – | – | – | 3.2 | 1.2 | <6.2 | – | – | – |
| 1741−038 | HPQ | 1.05 | 1.3 | 0.2 | 0.6 | 0.3 | <9.9 | – | – | – | 1 | 0.1 | – | – | – | – | 1.7 | 0.2 | 3.5 | 0.1 | 146 | 1 |
| 1749+096 | BLLac | 0.32 | 0.6 | 0.1 | 2.7 | 0.8 | 1.8 | 0.3 | 135 | 6 | 2.7 | 1.2 | – | – | – | – | 4.6 | 1.1 | 5.1 | 0.1 | 131 | 2 |
| 1823+568 | BLLac | 0.66 | – | – | – | – | – | – | – | – | – | – | – | – | – | – | 0.8 | 0.4 | 9.1 | 0.3 | 33 | 1 |
| 1827+062 | ? | – | 0.6 | 0.5 | – | – | – | – | – | – | 0.8 | 0.2 | – | – | – | – | 0.8 | 0.2 | 3.9 | 1.0 | 140 | 30 |
| 1923+210 | ? | – | – | – | – | – | – | – | – | – | – | – | – | – | – | – | 1.2 | 0.3 | 3.2 | 0.9 | 73 | 27 |
| 1928+738 | BLLac | 0.30 | 0.7 | 0.3 | 1.1 | 0.2 | – | – | – | – | 1.3 | 0.1 | – | – | – | – | 1.8 | 0.7 | 1.7 | 0.8 | 7 | 32 |
| 2005+642 | FSRQ | 1.57 | – | – | – | – | – | – | – | – | – | – | – | – | – | – | – | – | 3.0 | 0.1 | 157 | 1 |
| 2013+370 | BLLac | – | 0.5 | 0.1 | 2.2 | 0.7 | 4.9 | 0.2 | 150 | 1 | 2.5 | 0.3 | – | – | – | – | 3.0 | 0.5 | 2.9 | 0.1 | 38 | 1 |
| 2037+511 | QSO | 1.69 | 0.6 | 0.1 | 0.7 | 0.1 | – | – | – | – | 0.9 | 0.1 | – | – | – | – | 1.1 | 0.2 | 1.0 | 0.3 | 76 | 32 |

**Table 1.** continued.

| | | | | | 1.3mm band | | | | | | 2mm band | | | | | | 3mm band | | | | | |
|---|---|---|---|---|---|---|---|---|---|---|---|---|---|---|---|---|---|---|---|---|---|---|
| name (B1950) | type | $z$ [1] | $\alpha$ [1] | $\sigma_\alpha$ [1] | $S_\nu$ [Jy] | $\sigma_S$ [Jy] | $m_L$ [%] | $\sigma_m$ [%] | $\chi$ [°] | $\sigma_\chi$ [°] | $S_\nu$ [Jy] | $\sigma_S$ [Jy] | $m_L$ [%] | $\sigma_m$ [%] | $\chi$ [°] | $\sigma_\chi$ [°] | $S_\nu$ [Jy] | $\sigma_S$ [Jy] | $m_L$ [%] | $\sigma_m$ [%] | $\chi$ [°] | $\sigma_\chi$ [°] |
| 2047+039 | BLLac | – | – | – | – | – | – | – | – | – | – | – | – | – | – | – | 0.4 | 0.1 | 8.9 | 0.1 | 163 | 1 |
| 2131−021 | BLLac | 1.29 | – | – | – | – | – | – | – | – | – | – | – | – | – | – | 1.1 | 0.4 | 10.8 | 0.3 | 86 | 1 |
| 2134+004 | LPQ | 1.93 | 0.7 | 0.2 | – | – | – | – | – | – | 1.2 | 0.1 | – | – | – | – | 1.5 | 0.2 | 1.4 | 0.1 | 25 | 3 |
| 2201+315 | LPQ | 0.30 | – | – | – | – | – | – | – | – | – | – | – | – | – | – | 1.5 | 0.1 | 3.1 | 0.9 | 130 | 1 |
| 2223−052 | BLLac | 1.40 | 0.8 | 0.2 | 2.4 | 0.9 | 6.3 | 0.4 | 21 | 2 | 4.2 | 0.6 | – | – | – | – | 4.6 | 0.6 | – | – | – | – |
| 2251+158 | HPQ | 0.86 | 0.0 | 0.1 | 14.7 | 8.0 | – | – | – | – | 23.3 | 7.9 | – | – | – | – | 16.9 | 8.2 | 2.6 | 1.2 | 63 | 41 |
| 2344+092 | ? | – | – | – | – | – | – | – | – | – | – | – | – | – | – | – | 0.3 | 0.1 | 8.2 | 1.6 | 145 | 6 |

**Table 2.** Observations journal of our AGN survey. For each *source* we give the B1950 name (for some also a frequently used alternative name), the number of observations $N_{obs}$, and the number of polarization detections $N_{det}$. For each *observation* we give the observing date, the observing frequency $\nu$, a flag if polarization was detected, and the statistical accuracy of the degree of polarization $\delta m_L$ as quality indicator. Signs "–" denote "not applicable as no polarization was detected".

| name (B1950) | other name | $N_{obs}$ [1] | $N_{det}$ [1] | observing date | $\nu$ [GHz] | polarization detected? | $\delta m_L$ [%] |
|---|---|---|---|---|---|---|---|
| 0007+106 | | 2 | 1 | 24-AUG-2008 | 83 | no | – |
| | | | | 31-OCT-2009 | 154 | yes | 0.11 |
| 0007+171 | | 1 | 0 | 24-OCT-2009 | 93 | no | – |
| 0059+581 | | 5 | 5 | 07-MAY-2007 | 99 | yes | 0.06 |
| | | | | 10-MAY-2007 | 99 | yes | 0.05 |
| | | | | 10-AUG-2008 | 145 | yes | 0.20 |
| | | | | 01-APR-2009 | 110 | yes | 0.07 |
| | | | | 29-MAY-2009 | 90 | yes | 0.05 |
| 0106+013 | | 2 | 0 | 22-SEP-2007 | 100 | no | – |
| | | | | 21-DEC-2007 | 114 | no | – |
| 0119+115 | | 11 | 11 | 02-AUG-2007 | 93 | yes | 0.08 |
| | | | | 10-AUG-2007 | 93 | yes | 0.08 |
| | | | | 13-AUG-2007 | 93 | yes | 0.12 |
| | | | | 21-DEC-2007 | 114 | yes | 0.16 |
| | | | | 30-JUL-2008 | 83 | yes | 0.13 |
| | | | | 02-MAR-2009 | 170 | yes | 0.57 |
| | | | | 08-MAR-2009 | 170 | yes | 0.20 |
| | | | | 29-JUN-2009 | 83 | yes | 0.12 |
| | | | | 30-JUN-2009 | 83 | yes | 0.11 |
| | | | | 05-OCT-2009 | 83 | yes | 0.11 |
| | | | | 06-OCT-2009 | 83 | yes | 0.09 |
| 0122−003 | | 1 | 1 | 12-JUL-2007 | 106 | yes | 0.18 |
| 0212+735 | | 1 | 1 | 10-AUG-2008 | 145 | yes | 0.27 |
| 0215+015 | | 5 | 5 | 30-JUL-2009 | 95 | yes | 0.30 |
| | | | | 07-AUG-2009 | 104 | yes | 0.12 |
| | | | | 08-AUG-2009 | 104 | yes | 0.04 |
| | | | | 09-AUG-2009 | 104 | yes | 0.17 |
| | | | | 06-SEP-2009 | 212 | yes | 1.08 |
| 0221+067 | | 1 | 1 | 31-AUG-2009 | 98 | yes | 0.32 |
| 0234+285 | | 4 | 4 | 03-JUL-2007 | 102 | yes | 0.12 |
| | | | | 07-JUL-2007 | 102 | yes | 0.07 |
| | | | | 20-JUN-2008 | 115 | yes | 0.08 |
| | | | | 25-DEC-2008 | 115 | yes | 0.06 |
| 0235+164 | | 5 | 2 | 14-JUN-2007 | 113 | no | – |
| | | | | 30-JUN-2007 | 95 | no | – |
| | | | | 22-AUG-2008 | 89 | yes | 0.05 |
| | | | | 02-JAN-2009 | 95 | yes | 0.03 |
| | | | | 12-JUN-2009 | 115 | no | – |
| 0239+108 | | 2 | 0 | 04-JUL-2008 | 151 | no | – |
| | | | | 15-JUL-2008 | 129 | no | – |
| 0316+413 | 3C84 | 13 | 0 | 13-JUL-2007 | 90 | no | – |
| | | | | 04-AUG-2007 | 113 | no | – |
| | | | | 12-AUG-2007 | 113 | no | – |
| | | | | 19-SEP-2007 | 90 | no | – |
| | | | | 23-OCT-2007 | 227 | no | – |
| | | | | 24-OCT-2008 | 227 | no | – |
| | | | | 25-OCT-2008 | 227 | no | – |
| | | | | 24-NOV-2008 | 227 | no | – |
| | | | | 11-JUL-2009 | 113 | no | – |
| | | | | 20-JUL-2009 | 113 | no | – |
| | | | | 27-JUL-2009 | 81 | no | – |
| | | | | 26-OCT-2009 | 113 | no | – |
| | | | | 27-OCT-2009 | 113 | no | – |
| 0333+321 | | 13 | 3 | 01-APR-2008 | 93 | no | – |
| | | | | 02-DEC-2008 | 203 | no | – |
| | | | | 16-DEC-2008 | 110 | no | – |
| | | | | 25-DEC-2008 | 115 | no | – |
| | | | | 11-JAN-2009 | 203 | no | – |
| | | | | 13-JAN-2009 | 203 | no | – |
| | | | | 30-MAR-2009 | 87 | no | – |
| | | | | 03-JUL-2009 | 81 | no | – |
| | | | | 27-JUL-2009 | 81 | no | – |
| | | | | 06-AUG-2009 | 89 | no | – |
| | | | | 19-NOV-2009 | 89 | yes | 0.04 |
| | | | | 25-NOV-2009 | 89 | yes | 0.03 |
| | | | | 09-DEC-2009 | 203 | yes | 0.39 |
| 0336−019 | | 3 | 2 | 01-AUG-2008 | 134 | no | – |
| | | | | 22-AUG-2008 | 89 | yes | 0.08 |
| | | | | 24-FEB-2009 | 230 | yes | 0.99 |
| 0355+508 | NRAO150 | 7 | 7 | 26-OCT-2007 | 89 | yes | 0.03 |
| | | | | 28-OCT-2007 | 91 | yes | 0.03 |

**Table 2.** continued.

| name (B1950) | other name | $N_{\rm obs}$ [1] | $N_{\rm det}$ [1] | observing date | $\nu$ [GHz] | polarization detected? | $\delta m_L$ [%] |
|---|---|---|---|---|---|---|---|
| | | | | 11-OCT-2008 | 169 | yes | 0.11 |
| | | | | 08-APR-2009 | 93 | yes | 0.03 |
| | | | | 02-SEP-2009 | 113 | yes | 0.08 |
| | | | | 16-NOV-2009 | 113 | yes | 0.04 |
| | | | | 22-AUG-2008 | 86 | yes | 0.02 |
| 0400+258 | | 2 | 2 | 29-OCT-2009 | 89 | yes | 0.15 |
| | | | | 17-NOV-2009 | 89 | yes | 0.17 |
| 0415+379 | 3C111 | 21 | 14 | 18-MAR-2007 | 89 | yes | 0.14 |
| | | | | 25-MAR-2007 | 89 | no | – |
| | | | | 27-FEB-2008 | 89 | yes | 0.14 |
| | | | | 18-MAR-2008 | 146 | yes | 0.12 |
| | | | | 07-APR-2007 | 93 | no | – |
| | | | | 29-JUL-2007 | 86 | yes | 0.05 |
| | | | | 31-JUL-2007 | 86 | yes | 0.09 |
| | | | | 03-AUG-2007 | 84 | yes | 0.11 |
| | | | | 06-AUG-2007 | 87 | yes | 0.06 |
| | | | | 07-AUG-2007 | 84 | yes | 0.15 |
| | | | | 08-AUG-2007 | 84 | no | – |
| | | | | 11-AUG-2007 | 97 | yes | 0.07 |
| | | | | 15-AUG-2007 | 84 | yes | 0.08 |
| | | | | 18-AUG-2007 | 101 | yes | 0.08 |
| | | | | 31-AUG-2007 | 85 | no | – |
| | | | | 31-JAN-2008 | 84 | yes | 0.10 |
| | | | | 22-FEB-2008 | 110 | yes | 0.10 |
| | | | | 23-FEB-2008 | 220 | no | – |
| | | | | 05-APR-2008 | 93 | yes | 0.08 |
| | | | | 19-JUL-2008 | 115 | no | – |
| | | | | 08-SEP-2008 | 231 | no | – |
| 0420−014 | | 1 | 0 | 24-AUG-2008 | 115 | no | – |
| 0430+052 | | 2 | 2 | 21-JUL-2008 | 115 | yes | 0.20 |
| | | | | 14-NOV-2008 | 146 | yes | 0.27 |
| 0444+634 | | 1 | 1 | 08-APR-2009 | 93 | yes | 0.11 |
| 0446+112 | | 1 | 1 | 17-MAR-2009 | 90 | yes | 0.09 |
| 0507+179 | | 11 | 11 | 06-SEP-2007 | 85 | yes | 0.43 |
| | | | | 07-SEP-2007 | 98 | yes | 0.14 |
| | | | | 10-SEP-2007 | 86 | yes | 0.43 |
| | | | | 01-OCT-2007 | 86 | yes | 0.30 |
| | | | | 16-FEB-2008 | 231 | yes | 0.10 |
| | | | | 04-APR-2008 | 110 | yes | 0.02 |
| | | | | 05-APR-2008 | 93 | yes | 0.06 |
| | | | | 14-NOV-2008 | 146 | yes | 0.10 |
| | | | | 28-JAN-2009 | 231 | yes | 0.36 |
| | | | | 17-MAR-2009 | 90 | yes | 0.08 |
| | | | | 06-AUG-2009 | 89 | yes | 0.09 |
| 0528+134 | | 17 | 10 | 16-FEB-2007 | 231 | yes | 0.36 |
| | | | | 11-APR-2007 | 87 | yes | 0.09 |
| | | | | 15-APR-2007 | 115 | yes | 0.09 |
| | | | | 18-AUG-2007 | 101 | yes | 0.10 |
| | | | | 31-AUG-2007 | 85 | no | – |
| | | | | 01-SEP-2007 | 231 | no | – |
| | | | | 03-SEP-2007 | 115 | no | – |
| | | | | 22-SEP-2007 | 231 | yes | 0.31 |
| | | | | 03-OCT-2007 | 203 | yes | 0.22 |
| | | | | 12-NOV-2007 | 115 | yes | 0.35 |
| | | | | 15-FEB-2008 | 231 | no | – |
| | | | | 16-FEB-2008 | 231 | no | – |
| | | | | 18-MAR-2008 | 146 | yes | 0.08 |
| | | | | 23-SEP-2008 | 113 | yes | 0.07 |
| | | | | 17-NOV-2008 | 113 | no | – |
| | | | | 07-DEC-2008 | 231 | no | – |
| | | | | 24-DEC-2008 | 89 | yes | 0.13 |
| 0529+483 | | 1 | 1 | 16-NOV-2009 | 113 | yes | 0.06 |
| 0539−057 | | 1 | 1 | 17-DEC-2007 | 112 | yes | 0.01 |
| 0552+398 | | 3 | 1 | 03-SEP-2007 | 115 | no | – |
| | | | | 12-NOV-2007 | 115 | no | – |
| | | | | 22-APR-2009 | 153 | yes | 0.40 |
| 0611+131 | | 1 | 0 | 16-AUG-2009 | 89 | no | – |
| 0716+714 | | 4 | 4 | 21-AUG-2009 | 115 | yes | 0.11 |
| | | | | 09-NOV-2009 | 115 | yes | 0.09 |
| | | | | 15-NOV-2009 | 115 | yes | 0.20 |
| | | | | 30-DEC-2009 | 88 | yes | 0.09 |
| 0727−115 | | 1 | 0 | 03-DEC-2008 | 113 | no | – |
| 0748+126 | | 1 | 1 | 15-APR-2007 | 115 | yes | 0.13 |
| 0804+499 | | 1 | 1 | 25-SEP-2007 | 86 | yes | 0.05 |

**Table 2.** continued.

| name (B1950) | other name | $N_{obs}$ [1] | $N_{det}$ [1] | observing date | $\nu$ [GHz] | polarization detected? | $\delta m_L$ [%] |
|---|---|---|---|---|---|---|---|
| 0820+560 | | 3 | 2 | 01-OCT-2009 | 115 | no | – |
| | | | | 03-OCT-2009 | 115 | yes | 0.35 |
| | | | | 05-OCT-2009 | 115 | yes | 0.64 |
| 0833+585 | | 1 | 1 | 03-JUN-2009 | 101 | yes | 0.23 |
| 0836+710 | | 2 | 2 | 09-JAN-2007 | 94 | yes | 0.04 |
| | | | | 25-SEP-2007 | 86 | yes | 0.03 |
| 0851+202 | | 11 | 11 | 10-MAY-2007 | 85 | yes | 0.05 |
| | | | | 15-MAY-2007 | 102 | yes | 0.17 |
| | | | | 23-SEP-2007 | 97 | yes | 0.11 |
| | | | | 06-OCT-2007 | 97 | yes | 0.42 |
| | | | | 02-MAY-2008 | 97 | yes | 0.08 |
| | | | | 15-OCT-2008 | 231 | yes | 0.08 |
| | | | | 25-APR-2009 | 97 | yes | 0.05 |
| | | | | 27-NOV-2009 | 106 | yes | 0.27 |
| | | | | 02-DEC-2009 | 231 | yes | 0.37 |
| | | | | 05-DEC-2009 | 136 | yes | 0.25 |
| | | | | 27-DEC-2009 | 106 | yes | 0.48 |
| 0906+015 | | 6 | 1 | 02-DEC-2007 | 83 | no | – |
| | | | | 12-OCT-2008 | 103 | no | – |
| | | | | 29-NOV-2008 | 103 | no | – |
| | | | | 07-DEC-2008 | 134 | no | – |
| | | | | 04-JUN-2009 | 101 | yes | 0.10 |
| | | | | 14-OCT-2009 | 134 | no | – |
| 0923+392 | | 7 | 7 | 14-JAN-2007 | 115 | yes | 0.20 |
| | | | | 24-SEP-2007 | 108 | yes | 0.19 |
| | | | | 02-MAR-2008 | 159 | yes | 0.04 |
| | | | | 03-MAY-2008 | 115 | yes | 0.06 |
| | | | | 15-OCT-2008 | 231 | yes | 0.41 |
| | | | | 27-APR-2009 | 97 | yes | 0.29 |
| | | | | 30-APR-2009 | 231 | yes | 0.08 |
| 0954+556 | | 7 | 7 | 17-AUG-2009 | 95 | yes | 0.12 |
| | | | | 19-AUG-2009 | 95 | yes | 0.18 |
| | | | | 20-AUG-2009 | 95 | yes | 0.48 |
| | | | | 28-AUG-2009 | 95 | yes | 0.03 |
| | | | | 25-SEP-2009 | 115 | yes | 0.48 |
| | | | | 27-SEP-2009 | 115 | yes | 0.22 |
| | | | | 28-SEP-2009 | 115 | yes | 0.72 |
| 0954+658 | | 29 | 27 | 08-MAR-2007 | 235 | yes | 0.42 |
| | | | | 25-AUG-2007 | 85 | yes | 0.06 |
| | | | | 05-SEP-2007 | 85 | yes | 0.03 |
| | | | | 14-SEP-2007 | 84 | yes | 0.04 |
| | | | | 16-SEP-2007 | 84 | yes | 0.09 |
| | | | | 19-SEP-2007 | 84 | yes | 0.03 |
| | | | | 07-FEB-2008 | 208 | yes | 0.14 |
| | | | | 15-FEB-2008 | 207 | yes | 0.07 |
| | | | | 22-FEB-2008 | 208 | yes | 0.42 |
| | | | | 07-MAR-2008 | 208 | yes | 0.30 |
| | | | | 08-MAR-2008 | 208 | yes | 0.19 |
| | | | | 23-JUL-2008 | 94 | yes | 0.06 |
| | | | | 14-AUG-2008 | 165 | yes | 0.22 |
| | | | | 31-AUG-2008 | 91 | yes | 0.02 |
| | | | | 02-SEP-2008 | 88 | yes | 0.02 |
| | | | | 09-SEP-2008 | 100 | yes | 0.08 |
| | | | | 01-OCT-2008 | 111 | yes | 0.16 |
| | | | | 02-OCT-2008 | 92 | yes | 0.18 |
| | | | | 07-OCT-2008 | 154 | yes | 0.32 |
| | | | | 21-FEB-2009 | 137 | yes | 0.06 |
| | | | | 19-JUN-2009 | 101 | yes | 0.19 |
| | | | | 25-JUN-2009 | 101 | yes | 0.15 |
| | | | | 17-JUL-2009 | 101 | no | – |
| | | | | 22-JUL-2009 | 101 | no | – |
| | | | | 24-JUL-2009 | 101 | yes | 0.22 |
| | | | | 24-AUG-2009 | 115 | yes | 0.21 |
| | | | | 26-AUG-2009 | 115 | yes | 0.11 |
| | | | | 09-SEP-2009 | 115 | yes | 0.22 |
| | | | | 18-OCT-2009 | 150 | yes | 0.02 |
| 1005+066 | | 4 | 4 | 21-JAN-2008 | 83 | yes | 0.05 |
| | | | | 23-JAN-2008 | 83 | yes | 0.20 |
| | | | | 12-APR-2008 | 83 | yes | 0.18 |
| | | | | 08-JUL-2009 | 101 | yes | 0.21 |
| 1012+232 | | 2 | 0 | 19-NOV-2007 | 83 | no | – |
| | | | | 02-DEC-2007 | 83 | no | – |
| 1030+611 | | 10 | 7 | 16-SEP-2008 | 96 | yes | 0.08 |
| | | | | 15-FEB-2009 | 154 | yes | 0.10 |
| | | | | 16-FEB-2009 | 154 | yes | 0.19 |
| | | | | 17-MAY-2009 | 82 | yes | 0.08 |
| | | | | 20-MAY-2009 | 91 | yes | 0.15 |
| | | | | 09-JUN-2009 | 82 | no | – |
| | | | | 19-JUL-2009 | 109 | no | – |
| | | | | 30-AUG-2009 | 82 | no | – |
| | | | | 31-AUG-2009 | 82 | yes | 0.31 |

**Table 2.** continued.

| name (B1950) | other name | $N_{obs}$ [1] | $N_{det}$ [1] | observing date | $\nu$ [GHz] | polarization detected? | $\delta m_L$ [%] |
|---|---|---|---|---|---|---|---|
| | | | | 23-NOV-2009 | 82 | yes | 0.07 |
| 1040+244 | | 1 | 1 | 24-OCT-2008 | 115 | yes | 0.52 |
| 1044+719 | | 12 | 3 | 27-APR-2007 | 111 | no | – |
| | | | | 30-MAY-2007 | 113 | no | – |
| | | | | 18-JUN-2007 | 113 | no | – |
| | | | | 01-AUG-2007 | 94 | yes | 0.09 |
| | | | | 22-JUL-2008 | 97 | yes | 0.08 |
| | | | | 28-AUG-2008 | 160 | no | – |
| | | | | 29-AUG-2008 | 154 | no | – |
| | | | | 20-SEP-2008 | 154 | no | – |
| | | | | 20-MAY-2009 | 91 | no | – |
| | | | | 23-MAY-2009 | 91 | yes | 0.13 |
| | | | | 07-JUN-2009 | 151 | no | – |
| | | | | 11-JUN-2009 | 141 | no | – |
| 1055+018 | | 5 | 5 | 16-NOV-2007 | 115 | yes | 0.16 |
| | | | | 24-OCT-2008 | 115 | yes | 0.20 |
| | | | | 13-NOV-2008 | 102 | yes | 0.09 |
| | | | | 16-NOV-2008 | 260 | yes | 0.33 |
| | | | | 20-NOV-2008 | 102 | yes | 0.09 |
| 1144+402 | | 1 | 0 | 19-APR-2009 | 115 | no | – |
| 1150+497 | | 12 | 9 | 17-FEB-2007 | 109 | yes | 0.13 |
| | | | | 22-FEB-2007 | 109 | yes | 0.21 |
| | | | | 01-MAR-2007 | 109 | yes | 0.16 |
| | | | | 15-SEP-2007 | 86 | yes | 0.12 |
| | | | | 27-JAN-2008 | 94 | yes | 0.11 |
| | | | | 04-MAY-2008 | 91 | no | – |
| | | | | 05-MAY-2008 | 89 | yes | 0.11 |
| | | | | 20-APR-2009 | 115 | no | – |
| | | | | 23-APR-2009 | 115 | no | – |
| | | | | 12-JUN-2009 | 111 | yes | 0.08 |
| | | | | 14-AUG-2009 | 111 | yes | 0.07 |
| | | | | 20-DEC-2009 | 88 | yes | 0.14 |
| 1156+295 | | 6 | 4 | 23-JUN-2007 | 97 | no | – |
| | | | | 31-JUL-2007 | 102 | yes | 0.17 |
| | | | | 10-AUG-2007 | 115 | yes | 0.10 |
| | | | | 23-OCT-2008 | 115 | no | – |
| | | | | 13-MAY-2009 | 102 | yes | 0.06 |
| | | | | 19-MAY-2009 | 102 | yes | 0.05 |
| 1226+023 | 3C273 | 11 | 9 | 04-FEB-2007 | 222 | yes | 0.09 |
| | | | | 03-AUG-2007 | 115 | yes | 0.12 |
| | | | | 17-NOV-2007 | 115 | yes | 0.04 |
| | | | | 18-NOV-2007 | 231 | yes | 0.27 |
| | | | | 25-APR-2008 | 81 | yes | 0.08 |
| | | | | 08-JAN-2009 | 230 | no | – |
| | | | | 14-JAN-2009 | 230 | yes | 0.25 |
| | | | | 13-MAR-2009 | 230 | yes | 0.83 |
| | | | | 22-MAR-2009 | 230 | yes | 0.20 |
| | | | | 23-APR-2009 | 115 | no | – |
| | | | | 26-NOV-2009 | 86 | yes | 0.08 |
| 1253−055 | 3C279 | 1 | 1 | 02-MAR-2009 | 251 | yes | 0.34 |
| 1300+580 | | 1 | 1 | 12-JUN-2009 | 111 | yes | 0.46 |
| 1307+121 | | 1 | 1 | 13-AUG-2009 | 97 | yes | 0.11 |
| 1308+326 | | 10 | 9 | 21-FEB-2008 | 157 | yes | 0.14 |
| | | | | 07-MAY-2008 | 101 | yes | 0.05 |
| | | | | 28-MAY-2008 | 101 | yes | 0.05 |
| | | | | 17-AUG-2008 | 99 | yes | 0.36 |
| | | | | 19-AUG-2008 | 108 | yes | 0.07 |
| | | | | 21-AUG-2008 | 115 | yes | 0.37 |
| | | | | 28-SEP-2008 | 107 | yes | 0.09 |
| | | | | 27-OCT-2008 | 100 | yes | 0.04 |
| | | | | 08-NOV-2008 | 100 | no | – |
| | | | | 20-MAR-2009 | 239 | yes | 0.30 |
| 1334−127 | | 2 | 0 | 16-MAR-2007 | 212 | no | – |
| | | | | 02-MAR-2009 | 251 | no | – |
| 1345+125 | | 1 | 1 | 13-NOV-2009 | 97 | yes | 0.14 |
| 1354+195 | | 2 | 1 | 11-JUL-2007 | 102 | yes | 0.23 |
| | | | | 27-AUG-2007 | 102 | no | – |
| 1413+135 | | 3 | 0 | 30-JUL-2008 | 93 | no | – |
| | | | | 20-AUG-2008 | 93 | no | – |
| | | | | 08-SEP-2008 | 93 | no | – |
| 1417+385 | | 1 | 1 | 21-JAN-2007 | 107 | yes | 0.36 |
| 1418+546 | | 36 | 25 | 25-JAN-2008 | 94 | yes | 0.08 |
| | | | | 26-JAN-2008 | 91 | yes | 0.04 |
| | | | | 14-MAR-2008 | 94 | yes | 0.11 |

**Table 2.** continued.

| name (B1950) | other name | $N_{obs}$ [1] | $N_{det}$ [1] | observing date | $\nu$ [GHz] | polarization detected? | $\delta m_L$ [%] |
|---|---|---|---|---|---|---|---|
| | | | | 04-MAY-2008 | 96 | yes | 0.19 |
| | | | | 18-MAY-2008 | 163 | yes | 0.12 |
| | | | | 10-JUN-2008 | 137 | yes | 0.21 |
| | | | | 30-JUN-2008 | 93 | yes | 0.15 |
| | | | | 05-JUL-2008 | 101 | yes | 0.02 |
| | | | | 08-JUL-2008 | 154 | yes | 0.31 |
| | | | | 09-JUL-2008 | 154 | yes | 0.63 |
| | | | | 18-JUL-2008 | 112 | yes | 0.28 |
| | | | | 16-AUG-2008 | 158 | yes | 0.51 |
| | | | | 19-SEP-2008 | 108 | yes | 0.18 |
| | | | | 24-SEP-2008 | 95 | yes | 0.16 |
| | | | | 30-SEP-2008 | 160 | no | – |
| | | | | 19-OCT-2008 | 158 | no | – |
| | | | | 27-NOV-2008 | 158 | yes | 0.11 |
| | | | | 05-JAN-2009 | 91 | no | – |
| | | | | 14-JAN-2009 | 137 | yes | 0.14 |
| | | | | 30-JAN-2009 | 163 | yes | 0.15 |
| | | | | 14-FEB-2009 | 163 | yes | 0.08 |
| | | | | 12-MAR-2009 | 87 | no | – |
| | | | | 13-MAR-2009 | 163 | yes | 0.25 |
| | | | | 14-MAR-2009 | 160 | no | – |
| | | | | 01-APR-2009 | 155 | no | – |
| | | | | 05-APR-2009 | 165 | no | – |
| | | | | 28-MAY-2009 | 152 | no | – |
| | | | | 20-JUL-2009 | 109 | yes | 0.15 |
| | | | | 10-AUG-2009 | 95 | no | – |
| | | | | 22-AUG-2009 | 95 | no | – |
| | | | | 19-SEP-2009 | 107 | yes | 0.08 |
| | | | | 02-OCT-2009 | 115 | yes | 0.11 |
| | | | | 15-OCT-2009 | 253 | yes | 0.44 |
| | | | | 01-NOV-2009 | 253 | no | – |
| | | | | 17-NOV-2009 | 159 | yes | 0.17 |
| | | | | 26-DEC-2009 | 152 | yes | 0.07 |
| 1538+149 | | 2 | 2 | 26-AUG-2007 | 84 | yes | 0.34 |
| | | | | 18-AUG-2008 | 106 | yes | 0.12 |
| 1546+027 | | 1 | 1 | 16-MAR-2007 | 212 | yes | 0.34 |
| 1548+056 | | 2 | 2 | 02-JAN-2008 | 106 | yes | 0.11 |
| | | | | 28-JAN-2008 | 106 | yes | 0.16 |
| 1611+343 | | 6 | 6 | 14-AUG-2007 | 84 | yes | 0.25 |
| | | | | 02-JAN-2008 | 106 | yes | 0.03 |
| | | | | 28-JAN-2008 | 106 | yes | 0.09 |
| | | | | 25-DEC-2008 | 104 | yes | 0.29 |
| | | | | 29-MAR-2009 | 101 | yes | 0.08 |
| | | | | 06-AUG-2009 | 106 | yes | 0.03 |
| 1633+382 | | 1 | 0 | 04-APR-2008 | 262 | no | – |
| 1637+574 | | 24 | 16 | 13-FEB-2007 | 81 | yes | 0.08 |
| | | | | 06-MAR-2007 | 80 | no | – |
| | | | | 12-MAR-2007 | 104 | no | – |
| | | | | 18-MAR-2007 | 80 | yes | 0.01 |
| | | | | 29-AUG-2007 | 87 | yes | 0.10 |
| | | | | 01-SEP-2007 | 89 | yes | 0.01 |
| | | | | 06-OCT-2007 | 102 | yes | 0.18 |
| | | | | 26-OCT-2007 | 87 | yes | 0.04 |
| | | | | 09-APR-2008 | 110 | no | – |
| | | | | 23-APR-2008 | 110 | yes | 0.03 |
| | | | | 24-APR-2008 | 111 | yes | 0.06 |
| | | | | 27-APR-2008 | 100 | yes | 0.02 |
| | | | | 04-MAY-2008 | 90 | yes | 0.06 |
| | | | | 06-JUN-2008 | 104 | yes | 0.02 |
| | | | | 08-JUN-2008 | 104 | yes | 0.08 |
| | | | | 13-JUN-2008 | 108 | yes | 0.03 |
| | | | | 19-JUN-2008 | 108 | yes | 0.16 |
| | | | | 07-JUL-2008 | 96 | yes | 0.06 |
| | | | | 23-JUL-2008 | 115 | yes | 0.08 |
| | | | | 24-JUL-2008 | 115 | no | – |
| | | | | 20-SEP-2008 | 115 | no | – |
| | | | | 25-APR-2009 | 108 | no | – |
| | | | | 28-APR-2009 | 105 | no | – |
| | | | | 13-MAY-2009 | 92 | no | – |
| 1641+399 | 3C345 | 9 | 7 | 05-FEB-2007 | 234 | yes | 0.41 |
| | | | | 09-FEB-2007 | 234 | yes | 0.46 |
| | | | | 06-MAY-2007 | 108 | yes | 0.20 |
| | | | | 11-JUN-2008 | 143 | yes | 0.07 |
| | | | | 15-SEP-2008 | 170 | yes | 1.13 |
| | | | | 13-APR-2009 | 239 | no | – |
| | | | | 13-APR-2009 | 234 | no | – |
| | | | | 21-APR-2009 | 239 | yes | 0.26 |
| | | | | 19-NOV-2009 | 239 | yes | 0.32 |
| 1642+690 | | 7 | 7 | 18-JUL-2007 | 102 | yes | 0.15 |
| | | | | 24-FEB-2008 | 164 | yes | 0.18 |
| | | | | 27-FEB-2008 | 164 | yes | 0.08 |
| | | | | 18-MAR-2008 | 212 | yes | 0.09 |
| | | | | 03-MAY-2009 | 103 | yes | 0.02 |

**Table 2.** continued.

| name (B1950) | other name | $N_{obs}$ [1] | $N_{det}$ [1] | observing date | $\nu$ [GHz] | polarization detected? | $\delta m_L$ [%] |
|---|---|---|---|---|---|---|---|
| | | | | 24-MAY-2009 | 104 | yes | 0.03 |
| | | | | 11-JUL-2009 | 105 | yes | 0.06 |
| 1655+077 | | 1 | 1 | 20-FEB-2008 | 85 | yes | 0.11 |
| 1730−130 | NRAO530 | 1 | 0 | 18-FEB-2008 | 115 | no | – |
| 1741−038 | | 3 | 1 | 16-APR-2007 | 89 | yes | 0.02 |
| | | | | 29-JAN-2009 | 231 | no | – |
| | | | | 21-JUN-2009 | 115 | no | – |
| 1749+096 | | 3 | 3 | 20-FEB-2007 | 225 | yes | 0.29 |
| | | | | 06-APR-2007 | 87 | yes | 0.04 |
| | | | | 16-APR-2007 | 89 | yes | 0.04 |
| 1823+568 | | 2 | 2 | 27-JUN-2007 | 100 | yes | 0.10 |
| | | | | 28-AUG-2007 | 89 | yes | 0.16 |
| 1827+062 | | 8 | 7 | 16-AUG-2007 | 93 | yes | 0.10 |
| | | | | 01-JUL-2008 | 104 | yes | 0.33 |
| | | | | 03-JUL-2008 | 114 | no | – |
| | | | | 14-NOV-2008 | 114 | yes | 0.40 |
| | | | | 06-OCT-2009 | 93 | yes | 0.03 |
| | | | | 01-NOV-2009 | 93 | yes | 0.12 |
| | | | | 23-NOV-2009 | 93 | yes | 0.23 |
| | | | | 27-NOV-2009 | 93 | yes | 0.21 |
| 1923+210 | | 2 | 2 | 16-AUG-2007 | 93 | yes | 0.05 |
| | | | | 04-APR-2008 | 93 | yes | 0.15 |
| 1928+738 | | 4 | 4 | 18-JUL-2007 | 102 | yes | 0.13 |
| | | | | 07-JUL-2009 | 93 | yes | 0.05 |
| | | | | 12-JUL-2009 | 85 | yes | 0.04 |
| | | | | 13-NOV-2009 | 93 | yes | 0.03 |
| 2005+642 | | 1 | 1 | 08-NOV-2009 | 85 | yes | 0.14 |
| 2013+370 | | 5 | 2 | 19-FEB-2007 | 81 | yes | 0.12 |
| | | | | 29-JAN-2008 | 220 | no | – |
| | | | | 16-FEB-2008 | 220 | yes | 0.21 |
| | | | | 15-MAR-2008 | 220 | no | – |
| | | | | 19-MAR-2009 | 93 | no | – |
| 2037+511 | | 6 | 3 | 19-MAR-2009 | 93 | yes | 0.07 |
| | | | | 14-JUN-2009 | 93 | no | – |
| | | | | 16-JUN-2009 | 93 | no | – |
| | | | | 12-JUL-2009 | 85 | no | – |
| | | | | 28-OCT-2009 | 85 | yes | 0.02 |
| | | | | 08-NOV-2009 | 85 | yes | 0.07 |
| 2047+039 | | 1 | 1 | 22-MAY-2008 | 92 | yes | 0.13 |
| 2131−021 | | 1 | 1 | 23-MAY-2009 | 104 | yes | 0.27 |
| 2134+004 | | 5 | 1 | 19-JUL-2007 | 98 | no | – |
| | | | | 03-AUG-2007 | 97 | no | – |
| | | | | 10-AUG-2007 | 97 | no | – |
| | | | | 06-SEP-2007 | 98 | yes | 0.07 |
| | | | | 23-MAY-2009 | 104 | no | – |
| 2201+315 | | 2 | 2 | 29-JUN-2007 | 115 | yes | 0.09 |
| | | | | 25-JUL-2008 | 105 | yes | 0.05 |
| 2223−052 | | 1 | 1 | 07-SEP-2008 | 267 | yes | 0.44 |
| 2251+158 | 3C454.3 | 5 | 5 | 12-AUG-2007 | 93 | yes | 0.10 |
| | | | | 06-MAY-2008 | 110 | yes | 0.10 |
| | | | | 08-MAY-2008 | 110 | yes | 0.20 |
| | | | | 26-DEC-2008 | 108 | yes | 0.12 |
| | | | | 27-DEC-2008 | 109 | yes | 0.14 |
| 2344+092 | | 5 | 5 | 27-DEC-2008 | 109 | yes | 0.13 |
| | | | | 05-MAY-2009 | 109 | yes | 0.19 |
| | | | | 11-NOV-2009 | 107 | yes | 0.15 |
| | | | | 18-NOV-2009 | 107 | yes | 0.20 |
| | | | | 21-NOV-2009 | 111 | yes | 0.46 |